\documentclass[aps,pra,twocolumn,superscriptaddress,amsmath,amssymb,showpacs]{revtex4-2}
\pdfoutput=1
\usepackage{algorithm}
\usepackage{color}
\usepackage{algorithmic}
\usepackage{amsmath}
\usepackage{amssymb}
\usepackage{amsthm}
\usepackage{bm}
\usepackage{thmtools}

\usepackage{array}
\usepackage{url}
\usepackage{hyperref}
\usepackage{graphicx}
\usepackage{bibentry}
\usepackage{microtype}
\usepackage{amsmath,amssymb,graphicx}
\usepackage{hyperref}
\usepackage{longtable}
\usepackage{amsfonts}
\usepackage[usenames,dvipsnames]{xcolor}

\definecolor{darkerblue}{rgb}{0.2,0.2,0.5}

\providecommand{\U}[1]{\protect\rule{.1in}{.1in}}

\newtheorem*{theorem*}{Theorem}

\newcommand{\ket}[1]{\left| #1 \right>} % for Dirac bras
\newcommand{\bear}{\begin{array}}
\newcommand{\ear}{\end{array}}

\newcommand{\beq}{\begin{eqnarray}}
\newcommand{\eeq}{\end{eqnarray}}
\newcommand{\beqa}{\begin{eqnarray}}
\newcommand{\eeqa}{\end{eqnarray}}

\newcommand{\nn}{\nonumber}

\newcommand{\bk}[1]{\left(#1\right)}
\newcommand{\Bk}[1]{\left[#1\right]}

\newcommand{\trace}{\operatorname{tr}}

\def\be{\begin{equation}}
\def\ee{\end{equation}}
\def\bea{\begin{eqnarray}}
\def\eea{\end{eqnarray}}

\def\la{\langle}
\def\ra{\rangle}

\def\sL{\mathcal{L}}

\begin{document}
\title{Attaining quantum limited precision of localizing an object in passive imaging}
\author{Aqil Sajjad}
\email{aqilsajjad@optics.arizona.edu}
\affiliation{James C. Wyant College of Optical Sciences, University of Arizona, Tucson, AZ 85721}
\author{Michael R Grace}
\email{michaelgrace@email.arizona.edu}
\affiliation{James C. Wyant College of Optical Sciences, University of Arizona, Tucson, AZ 85721}
\author{Quntao Zhuang}
\affiliation{Department of Electrical and Computer Engineering, University of Arizona, Tucson, AZ 85721}
\affiliation{James C. Wyant College of Optical Sciences, University of Arizona, Tucson, AZ 85721}
\author{Saikat Guha}
\email{saikat@arizona.edu}
\affiliation{James C. Wyant College of Optical Sciences, University of Arizona, Tucson, AZ 85721}
\affiliation{Department of Electrical and Computer Engineering, University of Arizona, Tucson, AZ 85721}

%\cite{[{Supplementary material: Information on derivations. [URL]}] SM}
\begin{abstract}
	We investigate our ability to determine the mean position, or {\em centroid}, of a linear array of equally-bright incoherent point sources of light, whose continuum limit is the problem of estimating the center of a uniformly-radiating object. We consider two receivers: an image-plane ideal direct-detection imager and a receiver that employs Hermite-Gaussian (HG) Spatial-mode Demultiplexing (SPADE) in the image plane, prior to shot-noise-limited photon detection. We compare the Fisher Information (FI) for estimating the centroid achieved by these two receivers, which quantifies the information-accrual rate per photon, and compare those with the Quantum Fisher Information (QFI): the maximum attainable FI by any choice of measurement on the collected light allowed by physics. We find that focal-plane direct imaging is strictly sub-optimal, although not by a large margin. We also find that the HG mode sorter, which is the optimal measurement for estimating the separation between point sources (or the length of a line object) is not only suboptimal, but it performs worse than direct imaging. We study the scaling behavior of the QFI and direct imaging's FI for a continuous, uniformly-bright object in terms of its length, and find that both are inversely proportional to the object's length when it is sufficiently larger than the Rayleigh length. Finally, we propose a two-stage adaptive modal receiver design that attains the QFI for centroid estimation.
\end{abstract}
\maketitle

\section{Introduction}

Rayleigh's criterion for the resolution of two incoherent light sources~\cite{LordRayleigh1879} remains one of the most important results in optical imaging. Based on diffraction effects, it tells us that we cannot resolve two objects whose angular separation is less than $\lambda/D$, where $\lambda$ is the wavelength and $D$ is the size of the receiver's aperture. This relies on a somewhat heuristic argument. A more rigorous estimate for the maximum estimation precision for the separation between the two point-like sources, when imaged by an ideal direct-detection focal plane array, can be obtained from the classical Cram\'er-Rao bound \cite{VanTrees2013}, expressed in terms of the Fisher Information (FI). This analysis still shows that the FI, whose inverse gives a lower bound on the variance of any unbiased estimator, and therefore serves as a measure for precision, sharply drops and approaches zero as the separation between the two sources falls below $\lambda/D$, approaching zero. In qualitative terms, both Rayleigh's criterion and the Cram\'er-Rao bound are essentially telling us the same common sense thing: if two objects are too close to each other, then it is hard to tell them apart, and hence determine their separation, if their {\em images}---blurred by the point-spread function (PSF) of the aperture---overlap so much that it is hard to tell if the image is that of a single point source or that of two closely-spaced sources. This also translates into our inability to resolve any features of the object that are too small compared to $\lambda/D$. Obtaining relevant information about objects in the sub-Rayleigh regime has therefore been a major topic of interest for a wide variety of fields ranging from astronomy~\cite{Mari2012-ps,Xu2018-li} to biological imaging~\cite{Rust2006-mf,Vicidomini2018-tm}.

\subsection{Super-resolution imaging using pre-detection mode sorting}

Recent findings based on quantum estimation theory show that it is possible to build new imaging devices that surpass Rayleigh's limit. One useful tool in quantum estimation theory is the quantum Cram\'er-Rao bound (QCRB), introduced by Helstrom~\cite{Helstrom1976}, expressed in terms of the quantum Fisher information (QFI): the QFI is an upper bound to the maximum FI attainable with any physically-allowed measurement scheme. Thus, if we find that the FI for a given measurement (the physical device that detects the information bearing light producing an electrical signal) is equal to the QFI, then we know that the said measurement is optimal, and no other measurement in its place will generate an estimate of the parameter of interest with a lower variance. The inverse statement also holds for estimating a single scalar parameter: if we find that there is a gap between the QFI, and the FI of a specific measurement, then it means that we can do better by employing some other measurement whose FI is equal to the QFI.

It turns out that for estimating the separation between two incoherent point sources, the QFI remains constant instead of shrinking to zero as the separation reduces to zero~\cite{Tsang2016b}, proving thereby that Rayleigh's criterion---in its commonly stated form---is an artifact of {\em direct}, i.e., intensity, detection in the image plane (corrupted by the fundamental Poisson shot noise), which discards valuable information in the phase of the field. For the case of a Gaussian PSF, Tsang {\em et al.} showed that the QFI-attaining (optimal) measurement can be realized by an image-plane Hermite-Gaussian (HG) spatial-mode demultiplexer (SPADE), followed by shot-noise-limited photon detection on those sorted modes. Along the same lines, Kerviche {\em et al.}~\cite{KGA17} and Rehacek {\em et al.}~\cite{Rehacek2017a} independently showed that the optimal measurement for a hard aperture pupil (sinc-function PSF) is realizable with an image-plane sinc-Bessel SPADE followed by photon detection. Moreover, for the same problem with an arbitrary aperture function, both these works propose measuring in a basis comprising the PSF and its orthogonalized derivatives, with the latter proving that such a measurement attains the quantum optimal performance provided the PSF is an even function. In~\cite{Zachary2019}, Dutton {\em et al.} generalized the work of Tsang {\em et al.} to the case of estimating the angular extent of $M > 1$ equidistant equally-bright point sources. In the limit where the number of points goes to infinity, this equates to the problem of estimating the length of a continuous line-shaped object with uniform brightness. It was shown that an image-plane HG SPADE is again the optimal measurement for estimating the length of such a uniform-brightness line object~\cite{Zachary2019}. In~\cite{Ang2016}, the results of Tsang {\em et al.} were extended to 2-dimensions, showing that the 2-d HG basis is quantum optimal. A 3-dimensional generalization was the subject of~\cite{Prasad2018} where the so-called Zernike basis functions were shown to attain the QFI. 

These mode sorters, however, need to be pointed exactly at the centroid of the incoherent point sources in order to obtain an FI for estimating the separation that equals the QFI. In fact, Tsang {\em et al.} showed that even a somewhat small misalignment of the SPADE can result in a large drop in its performance, especially in the regime where the separation is much smaller than the Rayleigh limit. In this limit, it is even possible to simultaneously estimate the separation and centroid of two (not necessarily equal brightness) point sources optimally; however, such a measurement likewise needs to be spatially aligned with respect to the intensity-weighted centroid of the sources in order to avoid a significant loss of performance \cite{rehacek2018}. This means that if the centroid is not perfectly known~\emph{a priori}, we first need to estimate it before we carry out the SPADE measurement to determine the separation. Based on this intuition, a two-stage optimization scheme was proposed by Grace {\em et al.} in~\cite{Grace2020c}. This receiver first estimates the centroid---a {\em nuisance parameter}---using direct detection, and once a good enough estimate has been obtained, the system then switches to the SPADE for finding the separation.
In addition to the above motivation for estimating the centroid first when we are mainly interested in estimating the separation, there is also the fact that finding the location of an object is an important problem in its own right. We can for instance be interested in locating a known object whose size and features we already know.

All the above-mentioned works on estimating the separation generally assume that the centroid can be determined fairly accurately from direct imaging. This intuition in part stems from the fact that Helstrom showed that for a single point source, ideal image-plane direct detection is the optimal measurement to estimate its position~\cite{Helstrom1976}. Moreover, Tsang {\em et al.} showed that the Fisher information for the centroid for two equally bright sources, while less than the QFI, is not suboptimal by a very large difference.

\subsection{Main results}

Since SPADE-like measurements are so sensitive to misalignment, even small improvements in estimating the centroid can be beneficial. With that in mind, in this paper, we present a thorough study of the optimal measurement to estimate the centroid for two or more equally bright incoherent point sources, assuming a Gaussian PSF, including the case of an infinite number of equally-bright equally-spaced point sources in a straight line of sub-Rayleigh length. The infinite case is naturally of special interest because we want to be able to estimate the position of continuum objects, such as localizing a star or planet (in astronomical imaging) or localizing a cellular structure (in biological imaging).

We investigate the performance of both direct imaging and the HG SPADE for centroid estimation. The interest in the latter arises from it being the optimal measure for the separation, and it is therefore worth studying whether it can also be useful for finding the centroid. This is also of interest from the perspective of a two-stage detection scheme such as the one proposed in~\cite{Grace2020c}, where first the centroid would be determined using direct imaging or some other more optimal measurement, and once a reasonable estimate has been made for it, the device would switch automatically to the HG SPADE measurement for finding the separation. We show that for two or more sources with equal separation, direct imaging is in fact {\em not} an optimal measurement for estimating the centroid. We also find that the HG SPADE is more sub-optimal than direct imaging. Moreover, the better the HG SPADE is aligned with the centroid, the worse its ability to determine the centroid. 

Finally, we present a two-stage adaptive modal measurement strategy that achieves the QFI for estimating the centroid of a constellation of $n$ equally-bright equally-spaced point sources. The strategy we present applies to finding the QFI-attaining receiver measurement for any $n$-point constellation.

\subsection{Organization of the paper}

In section 2, we introduce the overall set up for centroid estimation and the underlying assumptions of the model, give a brief overview of the classical and quantum Cram\'er-Rao bounds, and summarize the key findings of~\cite{Tsang2016b, Zachary2019} that are most relevant for our study of centroid estimation. In section 3, we present our results on centroid estimation for different numbers of equi-distant, uniformly bright emitters placed in a single line, including the infinite case of a uniformly bright object. We compare the performance of direct imaging and the HG SPADE with the QFI, and show that direct imaging is sub-optimal for centroid estimation, but not by a substantial amount. We then go on to discuss the performance of the HG SPADE, and show that it is mostly worse than direct imaging for locating an object, even though it gives us the QFI-attaining measurement for finding the size or end-to-end diameter. Finally, we present a two-stage measurement scheme that attains the QFI for centroid estimation in the limit of large integration time. In section 4, we present our conclusions. We also include several appendices at the end of the paper, proving important results and explaining known calculation methods but with additional details and clarifications that have been generally skipped in the literature and may be helpful for the reader.

\section{The physical set up}

For the set up and our basic assumptions about the physics, we closely follow the framework laid out in~\cite{Tsang2016b}, except that we generalize it to more than two incoherently-radiating point sources. For simplicity, we assume that the object and image planes are one-dimensional with unit magnification and that our light sources emit nearly monochromatic light with paraxial waves~\cite{LordRayleigh1879}. We also make the standard assumption---valid for optical-frequency radiation---that the average number of photons $\epsilon$ per temporal mode arriving at the image plane is much less than one, requiring many photons to be detected over a large number of temporal modes, to extract any useful information~\cite{Goo85Statistical, MW95, Labeyrie2006, Gottesman2012, Tsang2011}. 

\subsection{Quantum model for imaging scene made up of incoherently radiating point emitters}
Let $\lambda$ be the center wavelength, $W$ (measured in Hz) the spectral bandwidth of the collected light (around $\lambda$), and $T$ (measured in seconds) the integration time. In that time-bandwidth window, there are roughly $M \approx WT$ mutually-orthogonal temporal modes. We take $N = M\epsilon$ to be the mean number of photons received over the integration time, where $\epsilon$ is the mean number of photons collected per temporal mode. We now write the density operator of the photon field in a single temporal mode over the infinite number of mutually-orthogonal spatial modes spanning the receiver telescope's entrance pupil's spatial extent. In the (conventional) image plane, this density operator can be written as:
\be
\rho = (1-\epsilon) \rho_0
+\epsilon\rho_1
+O(\epsilon^2),
\ee
where $\rho_0 = |{\boldsymbol 0}\ra\la {\boldsymbol 0}|$ is the zero-photon or ``vacuum" state and $\rho_1$ is a single-photon quantum state, both of a single temporal mode over some infinite-basis of spatial modes. $O(\epsilon^2)$ denotes higher order terms in $\epsilon$, which we will ignore, since at visible frequencies, $\epsilon \ll 1$. The quantum state of all the collected light during the integration is given by $\rho^{\otimes M}$.

The one-photon state $\rho_1$ is a mixed state: an incoherent mixture of states $|\psi_s\rangle$, a pure state of one photon---spread over an infinite basis of spatial modes---of the image plane field, corresponding to the $s$-th point source making up the overall scene. For a scene comprised of $n$ equally bright point sources,
\be
\rho_1 =
\frac{1}{n}
\sum_{s=1}^n |\psi_s\ra\la\psi_s|,
\label{rho1_definition}
\ee
with,
\be
|\psi_s\ra = \int_{-\infty}^\infty dx \psi(x -x_s)|x\ra,
\label{psi_s-ket-definition}
\ee
where $\psi(x)$ is our coherent PSF, $x_s$ is the position of the $s$-th point source, and $|x\ra = {\hat a}^\dagger(x)|{\boldsymbol 0}\ra$ is the (un-physical) state of one photon localized exactly at the spatial point $x$ in the image plane, where the annihilation and creation operators obey the delta-function commutator: $\left[{\hat a}(x), {\hat a}^\dagger(x^\prime)\right] = \delta(x-x^\prime)$. We will consider a Gaussian PSF:
\be
\psi(x) =
\frac{1}{(2\pi\sigma^2)^{1/4}}
\exp\left[-x^2/(4\sigma^2)\right],
\label{PSF-definition}
\ee
where $\sigma = 1/(2\Delta k) = \lambda/(2\pi {\rm N A})$, with $\Delta k^2 \equiv \int_{-\infty}^{\infty} [\partial \psi(x)/\partial x]^2 dx$, $\lambda$ the center wavelength, and ${\rm N A}$ the effective numerical aperture.

Now, let us consider the model for {\em ideal direct imaging} in the traditional image plane: an infinite-size continuum active surface (i.e., infinitely small detector pixels with unity fill factor), where each of those pixels is a unity quantum efficiency shot-noise-limited photon-number-resolving detector with an infinite bandwidth, no read noise, and no dead time. This ideal continuum detector array generates a spatio-temporal photo-current process that---for the aforesaid model of collected light---is characterized by a space-time Poisson point process with a rate given by the squared-magnitude photon-unit field in the image plane. We remind the reader that, per our model, for each of the $M$ temporal modes, at most one photon can be detected, since $\epsilon \ll 1$. Given there is a photon in a particular temporal mode, the spatial probability density of its detection is given by:
\be
\Lambda(x)
= \frac{1}{n}
\sum_{s=1}^n
| \la x | \psi_s\ra |^2
= \frac{1}{n}\sum_{s=1}^n |\psi(x-x_s)|^2.
\label{Lambda-def2}
\ee
Therefore, over a single temporal mode, the expectation value of the number of photons being detected in a region of width $dx$ around $x$ is then given by a Poisson distribution with the mean of $\epsilon \Lambda(x)$~\cite{Ram2006, pawley2006, Labeyrie2006,Zmuidzinas2003CramrRaoSL}. Over $M$ temporal modes, the average number of photons detected (over the entire detector array) becomes $N = M\epsilon$, with an average of $N\Lambda(x)dx$ photons in a region of width $dx$ around $x$ in the image plane. 

In general, if we instead measure the received optical field by some other receiver (e.g., homodyne detection), which is associated with an observable $\hat{\mathcal Y}$, then the probability distribution of measurement outcomes is given by $P(\mathcal Y) = \la\mathcal Y | \rho_1 | \mathcal Y\ra$, where $\mathcal Y\in\mathbb{R}$ is a particular value of the observable and $\ket{\mathcal Y}$ is the eigenket associated with that value. 

\subsection{Quantum Fisher Information and the Cramer-Rao Bound}
Let us say we are presented with $N$ copies of a quantum state, i.e., $\rho^{\otimes N}$, and we wish to estimate a set of parameters $\left\{\theta_\mu\right\}$ embedded in $\rho$ by measuring $\hat{\mathcal Y}$ on each copy of $\rho$. In other words, we have the classical estimation theory problem, wherein we wish to estimate parameters $\left\{\theta_\mu\right\}$ embedded in a random variable $\mathcal Y$, by $N$ i.i.d. samples of $\mathcal Y$, each drawn from the distribution $P(\mathcal Y) = \la\mathcal Y | \rho | \mathcal Y\ra$. Consider a set of estimators $\tilde\theta_\mu(\mathcal Y)$ and the error covariance matrix:
\be
\Sigma_{\mu\nu} \equiv \int d\mathcal Y P(\mathcal Y)
\Bk{\tilde\theta_\mu(\mathcal Y)-\theta_\mu}
\Bk{\tilde\theta_\nu(\mathcal Y)-\theta_\nu}.
\ee
If $\tilde\theta_\mu(\mathcal Y)$ is an unbiased estimator, then it obeys the Cram\'er-Rao bound on the covariance matrix
\be
\Sigma \geq \mathcal J^{-1},
\label{Cramer-Rao-bound}
\ee
where for $N$ measurements,
\be
\mathcal J_{\mu\nu}
\equiv N \, \int d\mathcal Y
\frac{1}{P(\mathcal Y)}
\frac{\partial P(\mathcal Y)}{\partial\theta_\mu}
\frac{\partial P(\mathcal Y)}{\partial\theta_\nu}
\ee
is the Fisher Information matrix~\cite{VanTrees2013} associated with this specific chosen receiver measurement.
Moreover, if $\tilde\theta_\mu(\mathcal Y)$ is the maximum likelihood estimator, then we saturate the inequality in (\ref{Cramer-Rao-bound}) for large $N$.
The FI thus quantifies the performance of a measurement in determining the parameter we are interested in.
Therefore, ideally, we want to choose a measurement that maximizes the FI.

The quantum Cram\'er-Rao bound gives us the maximum possible FI that {\em any} physically-permissible measurement scheme could achieve. In other words,
\be
\Sigma \geq \mathcal J^{-1}
\geq \mathcal K^{-1},
\ee
where $\mathcal K$ is the quantum Fisher information (QFI) matrix~\cite{Helstrom1976}.
For $\rho^{\otimes N}$ encoding parameters of interest $\left\{\theta_\mu\right\}$, the QFI matrix is given by:
\be
\mathcal K_{\mu\nu}(\rho^{\otimes N})
\equiv N \trace\left(\rho \{\sL_\mu(\rho),\, \sL_\nu(\rho)\} \right),
\ee
where $\mathcal L_\mu(\rho)$ is the {\em symmetric logarithmic derivative} (SLD) of $\rho$ with respect to the parameter $\theta_\mu$.
It is a Hermitian operator that is defined by the relation:
\be
\frac{\partial\rho}{\partial\theta_\mu}
\!=\! \frac{1}{2}\left(\rho \, \sL_\mu(\rho) + \sL_\mu(\rho) \,\rho\right).
\label{SLD-relation}
\ee
If $\rho = \sum_j D_j |e_j\ra\la e_j|$ is the decomposition of $\rho$ in terms of its eigenvalues $D_j$ and eigenvectors $|e_j\ra$, then the SLD is given by:
\be
\sL_\mu(\rho) = \sum_{j,k; D_j + D_k \neq 0}
\frac{2}{D_j+D_k}
\la e_j | \frac{\partial\rho}{\partial\theta_\mu}| e_k\ra
| e_j\ra\la e_k|.
\label{sld}
\ee
Note that the QFI matrix does not depend on a particular choice of measurement, but is a property of the quantum state $\rho$. If the FI for a chosen measurement scheme is equal to the QFI, then we know that it is the best possible way to estimate the parameter(s) of interest.

\subsection{Estimating geometrical parameters of a linear point source constellation}
Where possible, it is convenient to work in terms of parameters that give a diagonal QFI and FI. For our physical system of a linear constellation of equi-distant uniformly bright light sources, it turns out that the QFI and the FI for direct imaging are both diagonal in terms of the {\em centroid}
\be
\theta_1 =\frac{\sum_{s=1}^n x_s}{n},
\label{centroid_def}
\ee
and the {\em separation} between the first and last point source
\be
\theta_2 = x_n - x_1.
\label{theta2_def}
\ee
In terms of these two parameters, the individual positions of the point sources are given as:
\be
x_s = \theta_1
-\frac{\theta_2}{2}
+\frac{(s-1)\theta_2}{n-1}, \;\; 1\leq s \leq n.
\label{x_s-definition}
\ee
This diagonality of the QFI and the direct imaging FI in terms of $\theta_1$ and $\theta_2$ arises from the symmetry of our physical set up around the centroid, and we prove this in Appendix \ref{FI-and-QFI-diagonality}. It is also worth noting that due to the physical symmetry around the centroid, the direct imaging FI and QFI matrices will be independent of $\theta_1$.

\begin{figure}[htb]
	\centering
	\includegraphics[width=\linewidth]{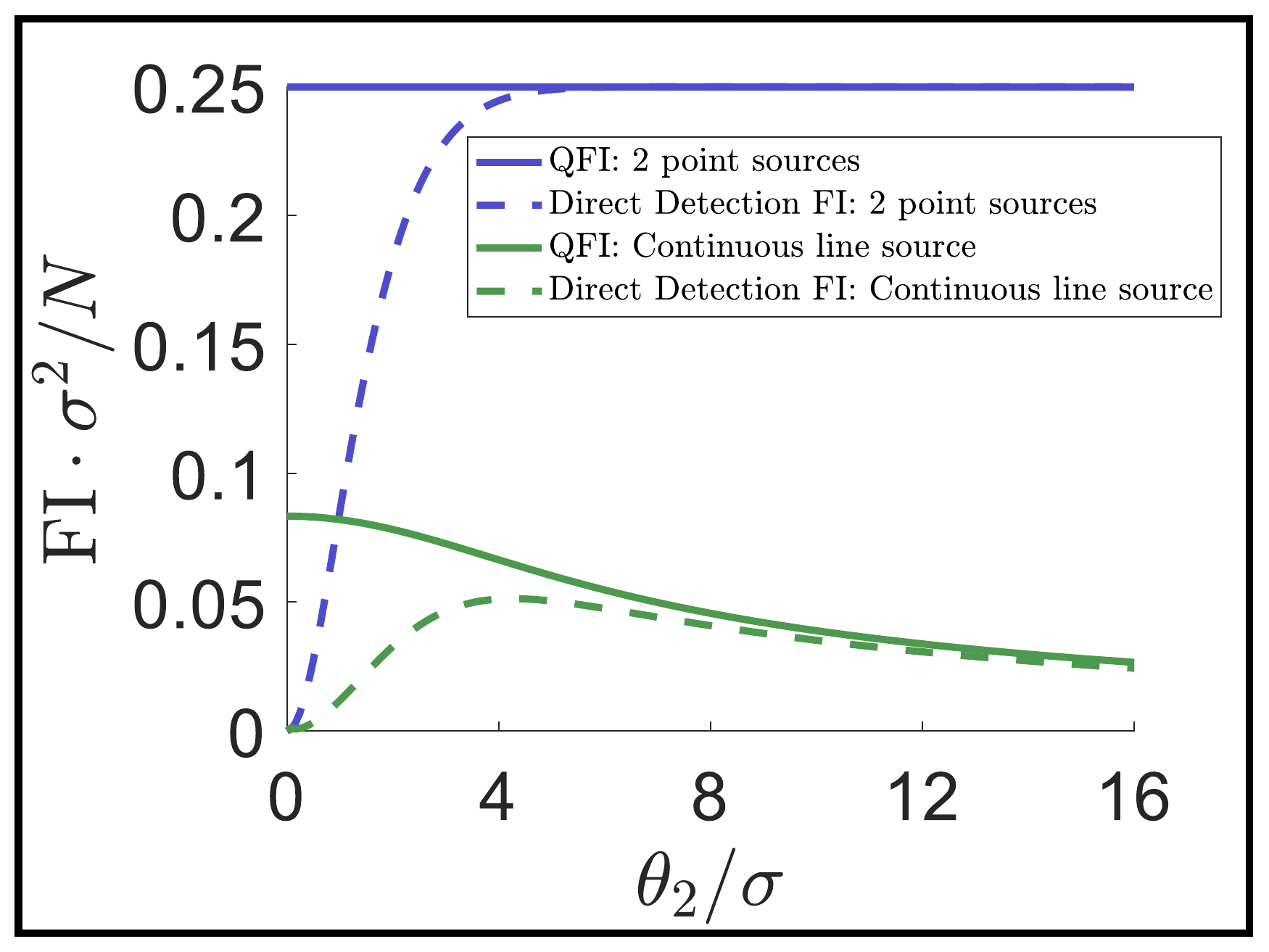}
	\caption{QFI (solid) and direct-imaging FI (dashed) for estimating the separation $\theta_2$ between $n$ point sources, plotted as a function of $\theta_2$. We show two sets of plots: one for $n=2$ point sources, and the other corresponding to $n \to \infty$, which corresponds to a continuous line-shaped object. A Gaussian PSF with width $\sigma$ is assumed.
	}
	\label{fig:1_Separation_plots}
\end{figure}

Tsang {\em et al.} studied the problem of estimating the separation between two equally bright point sources, i.e., $n=2$ in the above notation, assuming the centroid $\theta_1$ is known apriori~\cite{Tsang2016b}. The Fisher information for estimating the separation using direct imaging approaches zero as the separation goes to zero (see Fig.~\ref{fig:1_Separation_plots}). This is a manifestation of the so-called ``Rayleigh's curse", since the two sources become unresolvable when they are very close to each other, as their image-plane fields have a width comparable to their separation. However, Tsang {\em et al.} showed that the QFI for estimating $\theta_2$ is a non-zero constant even when the separation approaches zero: the same constant the direct-imaging FI approaches when $\theta_2 \to \infty$ (see Fig.~\ref{fig:1_Separation_plots}). This means that the so-called Rayleigh's curse is only an artifact of direct imaging rather than being a fundamental limit imposed by the PSF of the imaging system. They go on to show that if we carry out a measurement of the image-plane field using a Hermite Gaussian (HG) basis spatial-mode demultiplexer (SPADE) that is aligned perfectly with the centroid, and assuming that the centroid itself is perfectly known apriori, the FI attained by this measurement for estimating the separation $\theta_2$ equals the QFI, and hence being a quantum optimal measurement scheme for estimating the separation.

These results were generalized by Dutton {\em et al.} to the problem of finding the end-to-end separation for an arbitrary number of point sources ($n \ge 2$) in~\cite{Zachary2019}, including the $n \to \infty$ case of a line-shaped object of length $\theta_2$. They again found that the QFI does not fall to zero even as the separation becomes small, and that an HG SPADE aligned with the centroid attains the QFI.

We show these results in Figure \ref{fig:1_Separation_plots}, where we reproduce the plots reported in Tsang {\em et al.} for the QFI and the FI for direct imaging for finding the separation between 2 incoherent emitters~\cite{Tsang2016b}. We also show the same quantities for the continuous line source, reproducing the results of Dutton {\em et al.}~\cite{Zachary2019}. The QFI curves in both cases also represent the FI for the HG SPADE, since it attains the QFI. Additionally, we note that a ``binary SPADE" measurement---in which only the zeroth (or the first) image-plane HG mode is detected, by separating it from the rest of the light (which is also detected using a bucket detector)---attains the QFI in the limit of $\theta_2 \to 0$. This was shown for $n=2$ in~\cite{Tsang2016b,KGA17} and for $n \ge 2$ in~\cite{Zachary2019}.

It is however important to emphasize that all these results are strongly contingent on the alignment of the SPADE with the centroid. In fact, in the limit where $\theta_2 \to 0$ and $\theta_1 \gg \theta_2$, where the SPADE is aligned to the position $x=0$, the FI for the HG SPADE drops all the way from attaining the QFI to being $0$. This is for instance discussed in the Appendix D of~\cite{Tsang2016b} as well as in~\cite{Grace2020c}. 

Grace {\em et al.} studied the performance of a binary-SPADE measurement to estimate the separation of two point sources, when their centroid is not known apriori~\cite{Grace2020c}. They consider a two-stage adaptive receiver, where image-plane direct imaging is employed in the first segment of the optical integration time to obtain a estimate of the centroid, and the receiver then dynamically switches over to a second stage where a binary HG SPADE is employed whose center is aligned with respect of the (noisy) estimate of the centroid obtained from the first stage. Grace {\em et al.} developed an algorithm for that dynamic switching and the ensuing parameter estimation, which would enable $10$ to $100$ fold reduction in the integration time needed to obtain a desired (small) mean squared error in estimating $\theta_2$ despite no prior information of $\theta_1$ is assumed, compared to the scenario when image-plane direct detection is used for the entire integration time~\cite{Grace2020c}. 

The choice of image-plane direct detection to obtain a pre-estimate of the centroid during the first stage of the aforesaid adaptive receiver was driven by intuition. Image-plane direct imaging is quantum optimal (attains QFI) for localizing a single point source~\cite{Helstrom1976}, but suboptimal when it comes to estimating the centroid of two point sources~\cite{Tsang2016b}. Given the performance of SPADE-like measurements is extremely sensitive to misalignment, even small improvements in estimating the centroid can be very beneficial. This is our motivating reason to study the problem of centroid estimation. 

In this paper, we investigate the performance of both direct imaging and the HG SPADE for estimating the centroid $\theta_1$ of $n \ge 2$ point sources in a line spanning an angular length of $\theta_2$. The former because it is the simplest and the most obvious measurement, and the latter because it is worth asking if the SPADE can again outperform direct imaging in some region of parameter space for finding the centroid, just as it did for the separation. We also calculate the quantum limit (QFI) of centroid estimation to quantify the gaps to the FIs attained by the aforesaid two measurements. Finally, we describe an adaptive two-stage receiver design that would attain that QFI in the limit of long integration time.

\section{Quantum limit of localizing an object in passive imaging}

\subsection{The QFI and FI of direct imaging for a linear constellation of point sources}
\label{QFI-and-direct-imaging-behavior-section}

We now consider the behavior of the direct imaging FI and the QFI for estimating the centroid $\theta_1$ of a linear array of $n \ge 2$ equally-spaced point emitters spanning a total angular extent $\theta_2$. First, it is worth noting that the QFI and direct imaging FI should both be independent of the value of the centroid $\theta_1$ due to the assumption of a linear, shift-invariant physical imaging system. In the calculation of the FI for direct imaging, i.e., from the samples drawn from the spatial probability density $\Lambda(x)$ of photon clicks as in Eq.~\eqref{Lambda-def2}, this appears in the form of the shift symmetry of the variable of integration in~\eqref{direct-imaging-centroid-FI} from $x$ to $x-\theta_1$, which removes $\theta_1$ from the integrand.

In the case of a single light source, Helstrom showed that direct imaging is quantum optimal~\cite{Helstrom1976}. The QFI and direct imaging FI for this case can be calculated easily as we show in Appendix \ref{single-point-case-appendix}, and we find that they are both equal for any arbitrary PSF:
\be
\mathcal{K}_{\rm 1-pt} = \mathcal{J}_{\rm 1-pt}
= 4N\Delta k^2,
\ee
where
\be
\Delta k^2 \equiv \int_{-\infty}^\infty dx \left[\frac{\partial \psi(x)}{\partial x}\right]^2.
\label{Delta-k2-def}
\ee
For our Gaussian PSF defined in (\ref{PSF-definition}), this yields ${N}/{\sigma^2}$ for the QFI and the direct imaging FI. This is a result that we will regularly use throughout the rest of this paper since all the cases involving 2 or more points also have special limiting points where the FI and the QFI will approach this value.

For 2 point sources, the QFI has been worked out analytically by Tsang {\em et al.}~\cite{Tsang2016b}. We describe their calculation in Appendix \ref{2pt-QFI-calculation}, and simply state the result here. For the diagonal component of the QFI matrix involving the centroid, i.e., the QFI for estimating the centroid, they obtain:
\be
\mathcal{K}_{11} = 4N(\Delta k^2 -\gamma^2),
\label{K11-2pt-result}
\ee
where $\Delta k^2$ was defined in (\ref{Delta-k2-def}), and
\be
\gamma = \int_{-\infty}^\infty dx \, \frac{\partial \psi(x)}{\partial x} \, \psi(x -\theta_2).
\label{gamma-def}
\ee
It is worth noting that $\gamma$ goes to zero when $\theta_2$ goes to zero or infinity for any symmetric PSF. When $\theta_2$ goes to zero, the derivative of $\psi(x)$ is anti-symmetric, so the integral in (\ref{gamma-def}) tends to zero. On the other hand, when $\theta_2$ becomes large, then $\psi(x-\theta_2)$ and $\partial\psi(x)/\partial x$ overlap very little with each other for PSFs $\psi(x)$ that fall off to zero away from the origin. Therefore, again, $\gamma$ goes to zero. Consequently, the QFI approaches $4N\Delta k^2$ in these two limits, that is, the result for the single point-source case.
In between, however, there is a region where $\gamma$ is {\em not} zero, and we get a smaller QFI than that for a single point source. It is this regime where direct imaging is unable to attain the QFI for estimating the centroid.

\begin{figure}[htb]
	\centering
	\includegraphics[width=\linewidth]{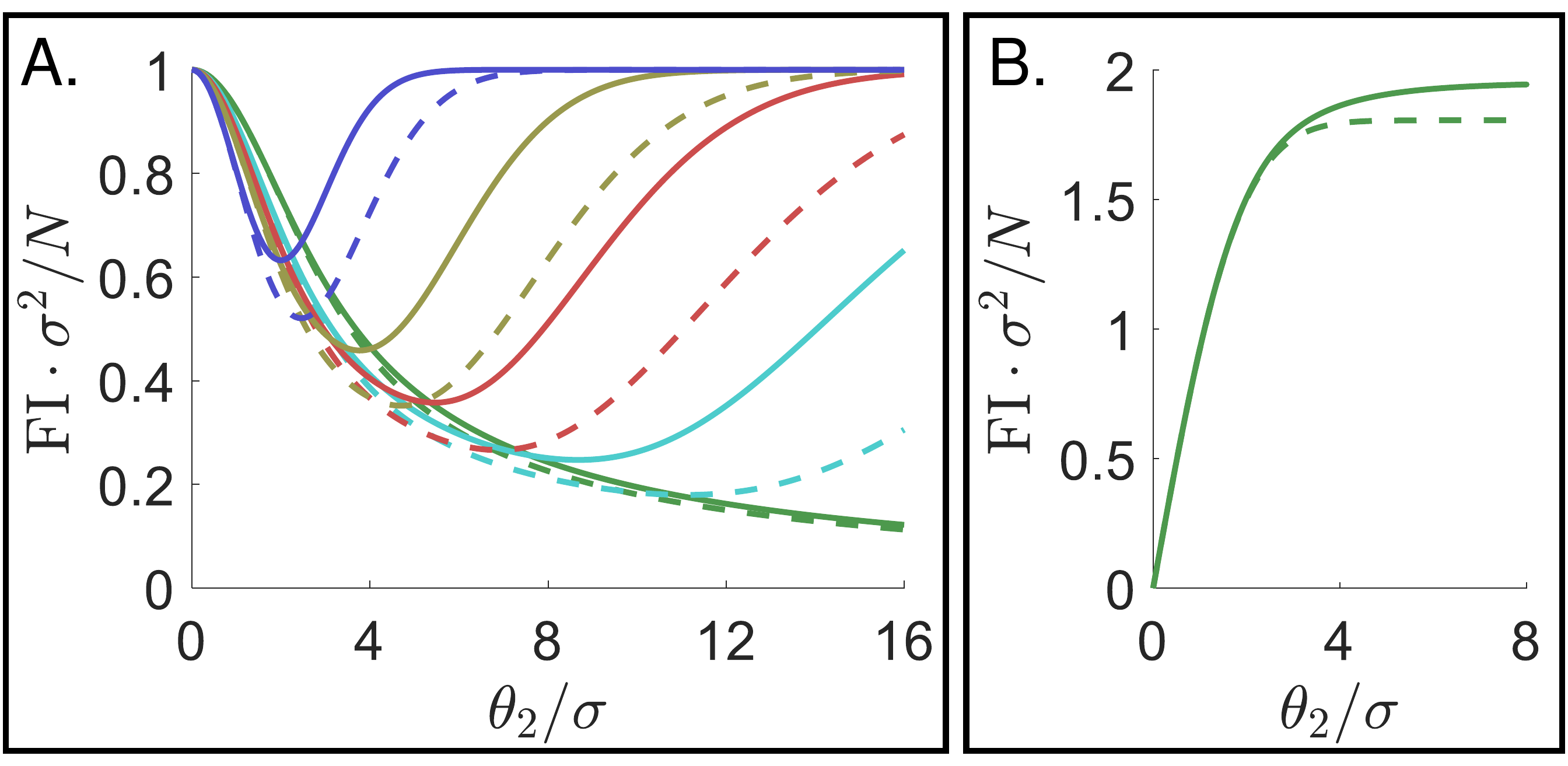}
	\caption{A. QFI (solid lines) and FI of direct imaging (dashed lines) for estimating the centroid of $n$ point sources (Blue: $n=2$; Gold: $n=3$; Red: $n=4$; Cyan $n=6$; Green: continuous line) with separation $\theta_2$ given a Gaussian PSF with width $\sigma$. B. QFI (solid line) and FI of direct imaging (dashed line) for a continuous line source with a constant photon flux per unit length of the source.
	}
	\label{fig:2_Direct imaging and QFI for diff n}
\end{figure}

The physical explanation for this behavior is that when the separation is very small, the impulse response of two point sources, i.e., their aperture-blurred fields in the image plane, each of width $\sigma$, look like the impulse response of a single point source at origin. Therefore, the problem of centroid estimation in this limit should reduce to that of finding the location of a single point source, for which image-plane direct detection is known to be quantum optimal~\cite{Helstrom1976}. For a slightly larger separation, the images of two point sources no longer overlap as much, and the effect of diffraction is to cause a decrease in the QFI. However, when the separation becomes significantly larger than the width of the PSF, the images of the two point sources fully separate with no overlap, in which regime their individual positions can be estimated separately---again quantum-optimally using image-plane direct detection---treating the two as single point sources. Each point source now emits half the light, and therefore the QFI for its location is $2N\Delta k^2$, but the total sum is still $4N\Delta k^2$. For the specific case of our Gaussian PSF
(\ref{PSF-definition}), we get
\be
\mathcal{K}_{11} =
\frac{N}{\sigma^2}
-\frac{N\theta_2^2}{4\sigma^4} \exp\left(-\frac{\theta_2^2}{4\sigma^2}\right),
\label{2pt-QFI}
\ee
which approaches the single point-source result of ${N}/{\sigma^2}$ in the $\theta_2\to 0$ and $\theta_2\to \infty$ limits with a dip in between as discussed above (see Fig.~\ref{fig:2_Direct imaging and QFI for diff n}A).

The FI of the centroid from direct imaging for the 2 point-source case has a somewhat similar qualitative behavior with the same physical intuition, except that its dip between the two limiting cases of $\theta \to 0$ and $\theta_2\to\infty$ is deeper. It is given by:
\be
\mathcal{J}_{11} = N \int dx \frac{1}{\Lambda(x)} \left(\frac{\partial \Lambda(x)}{\partial\theta_1}\right)^2,
\label{direct-imaging-centroid-FI}
\ee
where $\Lambda(x)$ is the probability density given in (\ref{Lambda-def2}).
We are unable to do this integral analytically, and therefore use numerical integration. The result was plotted along with the QFI in~\cite{Tsang2016b},
and we reproduce it in Fig. \ref{fig:2_Direct imaging and QFI for diff n} along with our results for when $n > 2$ point emitters constitute the scene. We see that the direct imaging FI approaches the QFI for small and large separation, as expected from the aforesaid intuitive explanation, for all $n \ge 2$. But, there is a gap in the region between these two limiting regimes. This gap is not too large, especially deep in the sub-Rayleigh regime. In particular, in the small $\theta_2$ regime for $n=2$, we can see this explicitly by Taylor expanding the QFI (\ref{2pt-QFI}) and also Taylor expanding the integrand of the direct imaging FI (\ref{direct-imaging-centroid-FI}) in the $\theta_2\to 0$ limit and integrating term by term. The resulting limiting behavior for the QFI and direct imaging FI are given by
\be
\mathcal{K}_{11} = \frac{N}{\sigma^2} - \frac{N\theta_2^2}{4\sigma^2} + \frac{N\theta_2^4}{16\sigma^6} - \frac{N\theta_2^6}{128\sigma^8} + O(\theta_2^8)
\ee
and
\be
\mathcal{J}_{11} = \frac{N}{\sigma^2} - \frac{N\theta_2^2}{4\sigma^2} + \frac{N\theta_2^4}{16\sigma^6} - \frac{N\theta_2^6}{64\sigma^8} + O(\theta_2^8)
\ee
from which we see that the two quantities vary only in $6^{\rm th}$ order in $\theta_2$. Tsang {\em et al.} argued~\cite{Tsang2016b} that we should be able to obtain a reasonable estimate for the centroid from direct imaging in order to correctly align the SPADE for estimating the separation, an intuition that was validated in the adaptive two-stage receiver designed and analyzed by Grace {\em et al.}~\cite{Grace2020c}.

Importantly, the above discussion using FI and QFI as performance benchmarking tools does not address the fact that if we are not in the regime $\theta_2/\sigma \ll 1$, the maximum likelihood estimator of $\theta_1$---either with direct detection or the quantum-optimal measurement as the receiver choice---would in general also depend on the true (apriori unknown) values of $\theta_1$ and $\theta_2$. The fact that with $\theta_1$ known apriori, the optimal measurement and estimator to estimate $\theta_2$ is the HG SPADE and is independent of the estimate of $\theta_2$, was a happy coincidence. 

We now come to generalizing the result in~\cite{Tsang2016b} for the 2-point-source case to a general number of incoherent point sources. For $n=3$ or more emitters, calculating the QFI becomes increasingly complicated as it involves diagonalizing larger and larger matrices. We therefore perform these diagonalizations numerically. The detailed procedure we employ for this purpose is described in Appendix \ref{HG-basis-QFI-procedure}, and here we focus on the results. Figure \ref{fig:2_Direct imaging and QFI for diff n}A shows the plots of QFI and direct imaging FI against the end-to-end separation $\theta_2$ for $n=2, 3, 4$ and $7$ emitters, as well as for a continuous line ($n \to \infty$). We see that as one would expect, both the QFI and direct imaging FI go to $N/\sigma^2$ when $\theta_2$ approaches zero, for any $n$ as well as the continuous line case.

As discussed above for $2$ sources, even for $n \ge 2$ sources, as $\theta_2$ increases from $0$, the QFI and direct imaging FI fall from $N/\sigma^2$ due to the diffraction-induced overlap among nearby point sources, and hence our ability to estimate the centroid decreases. However, the performance of direct imaging falls more rapidly than the QFI, and we see a small gap between the two. For any finite $n$, as $\theta_2$ increases to the extent that the $n$ point sources no longer significantly overlap, they essentially all become totally separate point sources, and their locations can be estimated individually as totally separate single emitters, just as we argued for $2$ points, in which regime the QFI is attainable with direct imaging. As a result, the QFI and direct imaging FI rise back towards the ${N}/{\sigma^2}$ value for a single emitter as $\theta_2$ becomes sufficiently large. However, as $n$ the number of point sources keeps increasing, $\theta_2$ must increase further for the points to become ``totally separate". Therefore we see that the QFI for $3$ sources has a minimum at a larger $\theta_2$ compared to that for the $2$ source case before it starts increasing again; and the direct imaging FI behaves the same way. Further increasing the number of point sources augments this effect, with QFI and direct imaging FI having their minima at even larger values of $\theta_2$ and requiring more and more separation for the QFI and FI to rise back toward the respective values for totally separated points.
For a continuous line source, i.e., $n = \infty$, since there is an infinite number of points next to each other, the QFI and direct imaging FI both monotonically decrease as we increase the length because the ``constituent point sources" comprising the uniformly-radiant object can never be ``totally separated". In this case, increasing $\theta_2$ only makes it more and more difficult to estimate the location of the centroid for a given total mean integrated photon number $N$.

In fact, it turns out that both the QFI and direct imaging FI for the continuous line scale as $1/\theta_2$ for large $\theta_2$. This can be seen as follows. Instead of assuming a fixed total number of photons $N$, let us consider the case when the total brightness of the object is proportional to its length. In other words, the total number of photons is $N\theta_2$, which amounts to simply multiplying the QFI and direct imaging FI values for $N$ photons, by $\theta_2$. These results are shown in Figure \ref{fig:2_Direct imaging and QFI for diff n}B as solid and dashed curves respectively. We see that they asymptote to constant values of about $1.95 N/\sigma^2$ and $1.80 N/\sigma^2$, for the QFI and direct imaging FI, respectively, which means that (1) the scaling behavior for constant total brightness (that does not scale with length) is indeed $1/\theta_2$, and (2) there is a constant-factor gap between direct imaging FI with QFI, in the large $\theta_2$ limit. It is also possible to see this scaling behavior analytically, even though we cannot carry out the full calculations for the QFI and direct imaging FI analytically and have to resort to numerical methods. We describe this in Appendix \ref{scaling-appendix}, where we outline the calculation for the continuum case where the sums over the infinite number of emitters, as in~\eqref{rho1_definition} and~\eqref{Lambda-def2} are replaced by integrals.

\begin{figure}[tb]
	\centering
	\includegraphics[width=\linewidth]{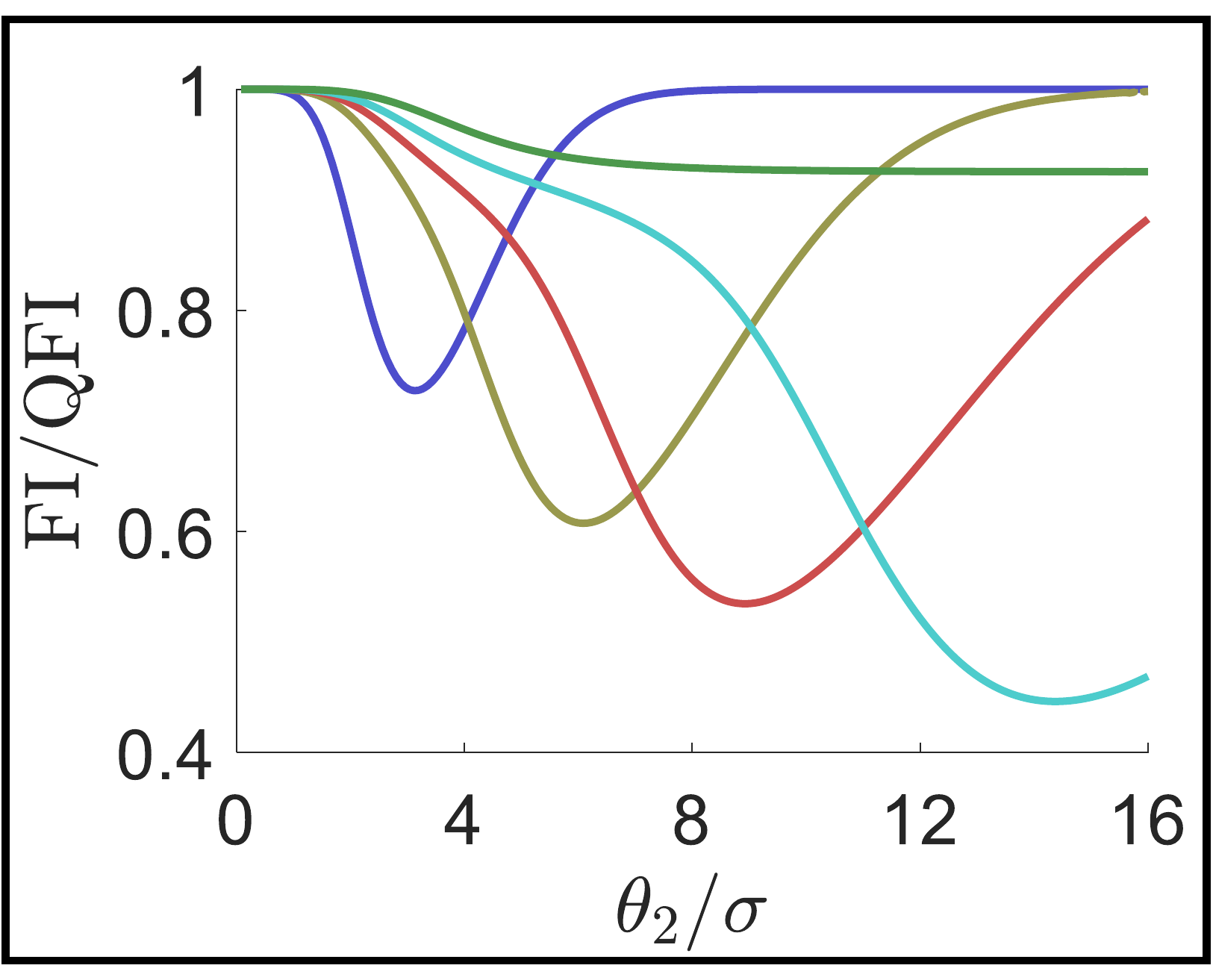}
	\caption{Ratio between the QFI and the direct imaging FI for estimating the centroid of $n$ point sources (Blue: $n=2$; Gold: $n=3$; Red: $n=4$; Cyan $n=6$; Green: continuous line) with end to end separation $\theta_2$ given a Gaussian PSF with width $\sigma$.
	}
	\label{fig:3_Direct imaging FI to QFI ratio for diff n}
\end{figure}

To get a clearer picture about how the performance of direct imaging compares with the quantum-optimal measurement for estimating the centroid, we consider the ratio of the direct imaging FI to the QFI. We plot this ratio against the end-to-end separation $\theta_2$ for different number of point sources as well as for the continuous line in Fig.~\ref{fig:3_Direct imaging FI to QFI ratio for diff n}. We see that overall the ratio is generally not too low, though in some places it goes down below 50 $\%$.
The minimum is about 73$\%$, $60\%$, $53\%$, and $45\%$ for $n=2, 3, 4$ and $6$ points, respectively. For the continuous line, the ratio continuously falls, asymptoting to a constant value of around $92.5\%$.

All these results mean that while direct imaging is sub-optimal, it is not significantly worse than the QFI. Therefore, it is not possible to get a significant improvement over direct imaging by using any other measurement. Yet, the fact that it falls down to about $72\%$ for $2$ point sources and even slightly below $50\%$ for a few more points, suggests that there is some room for improvement, especially in a situation where simply collecting more photons to get the same improvement is not the most desirable option. It is also worth mentioning that since the gap between direct imaging and QFI is generally small for the continuum case than for a small, finite number of emitters, direct imaging is closer to the optimal scheme for locating a full, uniformly bright object rather than one that has internal structure and/or sparsity. Finally, these results and our methods outlined above apply to any two- or three-dimensional point-source constellations, and it is possible that the gap between QFI and direct-imaging FI is higher for other more general constellations of point-emitters or continuous objects.

\subsection{Comparison with the HG SPADE's performance, and an interesting duality}
\label{SPADE-section}

We now consider an image-plane HG mode SPADE measurement~\cite{Yariv2006, Tsang2016b}, but for centroid estimation. Let us take the central position of this SPADE to be $x = 0$, such that the optical axis of the SPADE defines the 1D Cartesian coordinate system. The question we will consider is whether this can allow us to determine the centroid $\theta_1$ more efficiently than direct imaging. Let us begin with writing the quantum state of a single temporal mode of the collected image plane field with exactly one photon in the $q$-th HG mode. In other words, a single photon Fock state of the $q$-th HG mode is given by:
\be
|\phi_q\ra
= \int_{-\infty}^\infty dx \, \phi_q(x) \, |x\ra,
\ee
with $q = 0, 1, \ldots$, and
\be
\phi_q(x)
= 
\frac{(2\pi\sigma^2)^{-1/4}}{\sqrt{2^q q!}} H_q\left(\frac{x}{\sqrt{2}\sigma}\right)
\exp\left(-\frac{x^2}{4\sigma^2}\right),
\ee
where $H_q$ are the Hermite polynomials and $|x\rangle$ is the unphysical $1$-photon Fock state of the perfectly localized (delta-function) spatial mode at position $x$. Now, recalling (\ref{rho1_definition}), it is straightforward to see that if we pass the image-plane field through a HG mode sorter and detect photons on each mode, the probability of finding the photon in mode $q$ is:
\be
P(q)
= \frac{1}{n} \sum_{s=1}^n P_s(q),
\label{P-def}
\ee
where $P_s(q)$ is the probability for a photon from source $s$ to be found in mode $q$:
\be
P_s(q) \equiv |\la\phi_q|\psi_s\ra|^2
= \bigg|\int_{-\infty}^\infty dx \,\phi_q(x) \psi(x -x_s) \bigg|^2.
\ee
For our Gaussian PSF defined in (\ref{PSF-definition}), this gives
\be
P_s(q)
= \exp\left(-Q_s\right) \frac{Q_s ^q}{q!},
\label{P_s-def}
\ee
where
\be
Q_s = \frac{x_s ^2}{4\sigma^2}.
\label{Q_s-def}
\ee
It is now a straightforward exercise to obtain the Fisher information,
\be
\mathcal{J}_{{\rm HG}, 11}
= \sum_{q=0}^\infty \frac{N}{P(q)}
\left(\frac{\partial P(q)}{\partial \theta_1}\right)^2
\nonumber \ee\be
= \sum_{q=0}^\infty \,
\Bigg(
\sum_i Q_i (qQ_i ^{q-1} -Q_i ^q)^2 e^{-2Q_i}
\nn \ee\be
+ \sum_{i\neq j} 2 Q_{i j} (qQ_i ^{q-1} -Q_i ^q)(qQ_j ^{q-1} -Q_j ^q) e^{-Q_i -Q_j}
\Bigg)
\nn \ee\be
\times \frac{N}{n\sigma^2 q! \sum_i \exp\left(-Q_i\right) Q_i ^q},
\label{HG-SPADE-FI}
\ee
where $Q_{i j} \equiv ({x_i x_j})/({4\sigma^2}) = \sqrt{Q_i Q_j}\,$.

For the continuous line case of the number of points becoming infinite, we replace the sum in (\ref{P-def}) by an integral. We show in Appendix F that
\be
\mathcal{J}_{{\rm HG}, l, 11}
= N\sum_{q=0}^\infty \frac{\left(\exp\left(-Q_+\right) \frac{Q_+ ^q}{q!}
	-\exp\left(-Q_-\right) \frac{Q_- ^q}{q!} \right)^2}
{\theta_2 \int_{y =-\theta_2/2}^{\theta_2/2}
	\exp\left(-Q(y)\right) \frac{Q(y) ^q}{q!}dy}
\label{J_hg_line}
\ee
where
\be
Q_\pm = \frac{(\theta_1 \pm \theta_2/2)^2}{4\sigma^2}.
\ee
Note that this clearly has a dependence on $\theta_2$ other than the $1/\theta_2$ factor, and therefore does not obey the same type of scaling behavior for large $\theta_2$ that we found for the QFI and the direct imaging FI.

It is not clear how to do the sum over the FI contributions for the individual HG modes in (\ref{HG-SPADE-FI}) or (\ref{J_hg_line}) analytically, so we have to do this numerically. However, the series sum does simplify nicely for a few special cases:
\begin{enumerate}
	\item {\bf When $\theta_2$ approaches zero but $\theta_1$ does not}. This is essentially the limiting case where all the emitters effectively merge into a single one, but the SPADE is mis-aligned with the centroid by a constant amount. For this, the FI approaches the ${N}/{\sigma^2}$ value, the QFI for a single emitter as discussed in appendix \ref{HG-FI-single-point}. But then, for this $\theta_2\to 0$ case, we also know that the performance of direct imaging approaches the QFI~\cite{Helstrom1976}.
	
	\item {\bf When $\theta_1$ approaches zero but $\theta_2$ does not}. This is the case when the SPADE is almost perfectly aligned with the centroid, and the end-to-end separation is a constant. In this case, we get zero for the FI. This means, the HG SPADE yields tending-to-zero information about the centroid as the SPADE's alignment with the true centroid approaches near perfect.
	
	To see this, recall (\ref{P-def}), (\ref{P_s-def}) and (\ref{Q_s-def})
	and consider the partial derivative of $P(q)$:
	\be
	\frac{\partial P(q)}{\partial\theta_1}
	= \frac{1}{n}
	\sum_{s=1}^n \frac{\partial P_s(q)}{\partial\theta_1}
	\nn\ee\be
	= \sum_{s=1}^n \frac{x_s}{2n \sigma^2}
	\exp\left(-Q_s\right)
	\left(\frac{q Q_s ^{q-1}}{q!} -\frac{Q_s ^q}{q!}\right).
	\label{P-def-form}
	\ee
	Now, if we have an even number of emitters, then all their locations come in pairs of the form $x_s = \theta_1 \pm |c_s| \theta_2/2$,
	where $c_s$ is a constnat factor whose exact value depends on $s$.
	When $\theta_1 = 0$, these become $x_s = \pm |c_s|\theta_2/2$,
	and the corresponding $Q$ values (which are proportional to $x_s ^2$), are then equal for each pair.
	The sum in (\ref{P-def-form})
	then becomes zero since the contributions from the two points in each pair cancel due to the $x_s$ in front.
	If we have an odd number of sources, then all of them are in similar pairs, except the middle one at $x =\theta_1$. But this becomes zero for $\theta_1 = 0$, and therefore, we still get zero for the sum in (\ref{P-def-form}).
	
	This result has important implications for the two-stage set up of the kind being proposed in Ref.~\cite{Grace2020c}, in which we first obtain an estimate of the centroid from direct imaging in order to align an HG mode sorter in the second stage for estimating the separation or object size. Once we switch to the HG SPADE with a reasonably decent alignment close to the centroid, we will not be able to get any improvement in our estimate for the centroid.
	Our entire estimation precision of the centroid will therefore be based on the first measurement stage alone.
\end{enumerate}

\begin{figure}[htb]
	\centering
	\includegraphics[width=\linewidth]{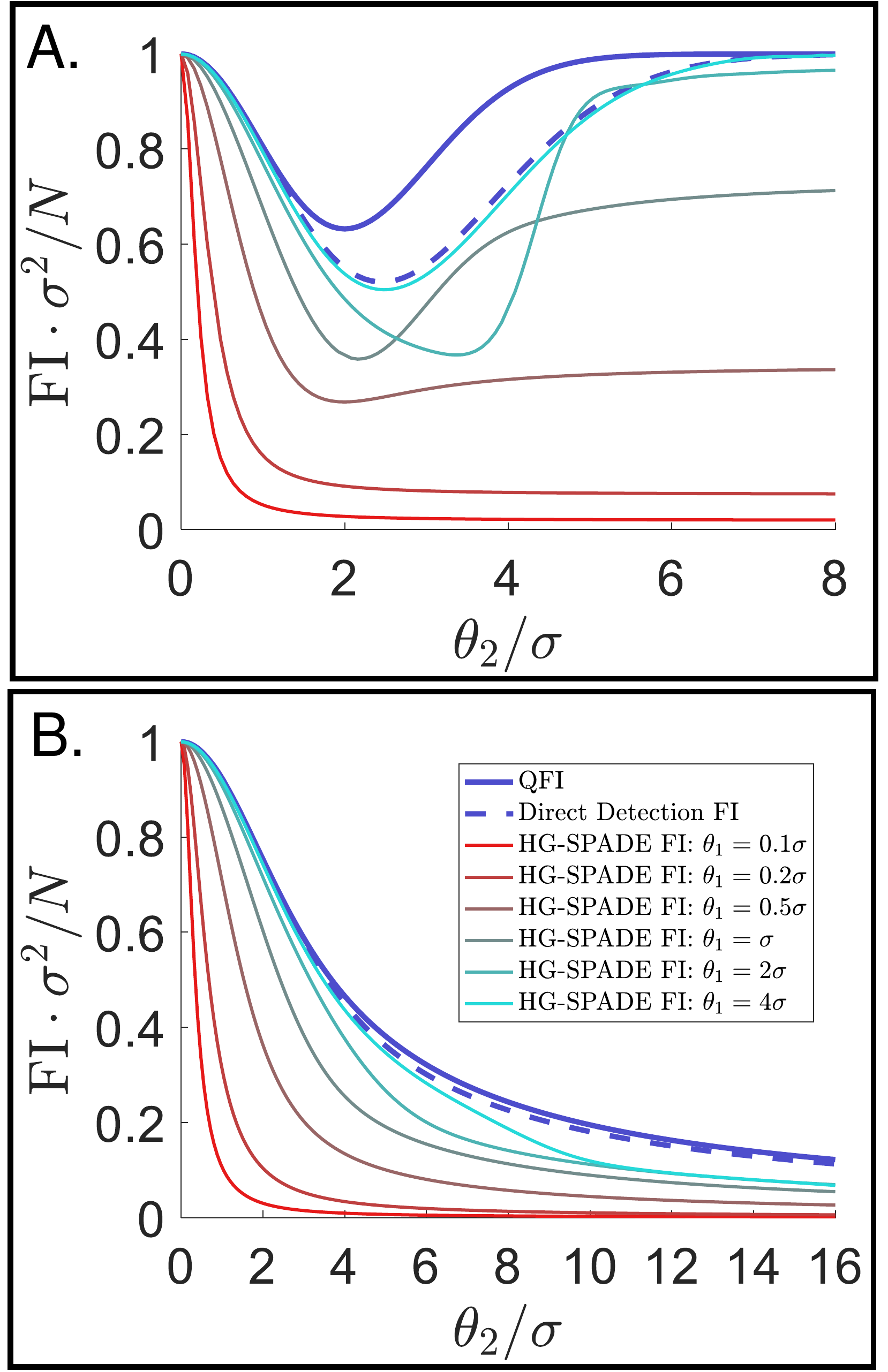}
	\caption{A. QFI, direct detection FI, and HG-SPADE FI for estimating the centroid of two point sources with separation $\theta_2$ and a centroid of $\theta_1$ from the origin (i.e., the SPADE alignment axis). B. Same as panel A, but for a continuous line source. 
	}
	\label{fig:4_HG SPADE performance}
\end{figure}

We now compare the performance of the HG SPADE with direct imaging and the QFI for centroid estimation. For this purpose, we focus on the 2-emitter and the continuum cases as our two examples that illustrate the overall pattern. For 2 sources, the plots of the QFI, direct imaging FI and the HG SPADE for different fixed values of the mis-alignment are given in Figure \ref{fig:4_HG SPADE performance}A. The same comparison for the continuum case is shown in Figure \ref{fig:4_HG SPADE performance}B. The QFI and direct imaging curves in these figures are of course the same as those shown in
Figure \ref{fig:2_Direct imaging and QFI for diff n}A,
and for the 2-emitter case, the QFI and direct imaging curves are the same as those shown in Ref.~\cite{Tsang2016b}, but that reference does not compare these with the HG SPADE's FI. We see that the performance of the HG SPADE is mostly worse than direct imaging for these graphs. It does however tend to converge with direct imaging from below in the $\theta_2 \to \infty$ limit. But it is never higher than the direct imaging curve, except in a very small region for the $\theta_1 = 2\sigma$ curve for the 2-emitter case. This is a special region between the $\theta_2\to 0$ and large $\theta_2$ extremes where the direct imaging FI drops sufficiently, and the FI for the HG SPADE rises enough to achieve a very small amount of superiority. It is worth noting that this is a region where the HG SPADE has enough mis-alignment with the centroid and is also pointed sufficiently away from either of the two emitters. We can also find some other such regions where this happens and the SPADE performs better than direct imaging, but the improvement is very small, and for most of the parameter space, the latter outperforms the former by a bigger margin. 

Therefore, the overall conclusion is that for all practical purpose, direct imaging is better than the HG SPADE for estimating the centroid, and that there is a small gap between the HG SPADE's performance and the ultimate quantum limit in the intermediary range of the object's length.

This highlights an interesting duality and complementarity between direct imaging and the HG SPADE. Unless the separation is large, direct imaging performs poorly for estimating the separation, especially in the sub-Rayleigh regime, whereas the HG SPADE, if aligned perfectly with the centroid, allows us to attain the QFI. But the SPADE generally does not perform very well for estimating the centroid, whereas direct imaging comes closer to attaining the QFI. We should however qualify this statement by pointing out that this complementarity is not totally perfect; it certainly holds when the SPADE points exactly at the centroid, but there are regions of the parameter space for non-zero alignment where this relationship is no longer true. The small region we saw in the 2-emitter case where the SPADE with a mis-alignment from the centroid of $2\sigma$ slightly outperforms direct imaging is an example of this. There is also a duality in the sense that a perfectly aligned SPADE attains the QFI for separation estimation but completely fails in determining the centroid, whereas a mis-aligned SPADE generally tends to be better for estimating the centroid rather than the separation.

\subsection{The quantum-optimal measurement scheme for estimating the centroid}

Having found that direct imaging is not optimal for centroid estimation in general, and the HG SPADE's performance is mostly worse than direct imaging, we now discuss a scheme for surpassing it and asymptotically attaining the QFI. In general, the projective quantum measurement given by the eigenbasis of the SLD for a given quantum state defines a physically allowed measurement that achieves the QFI~\cite{PhysRevLett.72.3439, Barndorff_Nielsen_2000, Paris2008}. This SLD measurement would translate to a SPADE (not the HG SPADE), followed by photon detection on the sorted spatial modes. This is because any projective measurement on a quantum state of one photon in many (spatial) modes---which is the case for the quantum description of the state of a single temporal mode of collected light in our problem---is always realizable by a passive linear optical transformation followed by photon detection~\footnote{The authors credit Saikat Guha and Ranjith Nair for realizing this well-known fact, in the context of this problem}.

However, since in this case the SLD depends on the true value of the centroid itself, this eigenbasis, and hence the aforesaid quantum-optimal SPADE to detect each temporal mode of the collected field, also depends upon the centroid. Therefore, we cannot carry out a measurement in this basis unless we already know the centroid, which is the variable we are trying to estimate in the first place. To get around this problem, we use a $2$-stage adaptive measurement scheme proposed by ~\cite{Barndorff_Nielsen_2000}, applied to our problem:
\begin{enumerate}
	\item Recall that $N = M\epsilon$ is the mean photon number of the total collected field, where $M$ is the number of temporal modes and $\epsilon \ll 1$ is the mean photon number per mode. The receiver's first stage uses a small proportion of the integration time, worth $N^\alpha$ mean photon number, with any $0 < \alpha < 1$, to obtain an initial maximum likelihood estimate $\hat \theta_1$ for the unknown centroid parameter $\theta_1$. The measurement used for this stage can be direct imaging, and does not have to be an optimal choice in any sense. The only requirement on this measurement is that it has a non-zero FI for estimating the centroid, $\theta_1$.
	
	\item Based on the estimate $\hat \theta_1$ obtained during the first stage, we carry out a measurement on each temporal mode of the remaining collected photons (of mean photon number $N-N^\alpha$) using the eigenbasis of the SLD of $\theta_1$, evaluated at $\hat \theta_1$. Based on this measurement, we obtain a maximum likelihood estimate for $\theta_1$.
\end{enumerate}

We would like to mention that using any other QFI attaining measurement in place of the SLD eigenbasis in the second stage will give the same performance in theory. Since the SLD basis is not necessarily the only measurement that attains the QFI, there may be other alternatives too; for example, the linear interferometric approach put forward by~\cite{Cosmo2020} is worth investigating for this purpose.

The above procedure prescribes a measurement that asymptotically reaches the efficiency of the QFI as $N$ tends to infinity. This two-stage scheme is also described in~\cite{2005atqs.book162H, PhysRevA.61.042312}
and section 6.4 of ~\cite{Hayashi2006}, with the specific choice of $\alpha = 1/2$, and with the condition that the FI for the first stage should be non-zero. 

While we refer the reader to the above references for detailed derivation of why this two-stage approach should attain the efficiency of the QFI in the large $N$ limit even though the SLD depends upon the \textit{a-priori}-unknown parameter, here is a short summary of the argument. If $N$ is sufficiently large, $N^\alpha$ or $\sqrt{N}$ in particular will also be large. And therefore, the variance of the estimate $\hat \theta_1$ scales as $1/(\mathcal{J}_1 N^\alpha)$, where $\mathcal{J}_1 > 0$ is the FI of this stage-one measurement, hence approaching zero as $N \to \infty$, around the true (\textit{a-priori}-unknown) value of $\theta_1$.
Now, the Fisher information in stage-two should be $(N-N^\alpha) \mathcal K$, if we measure in the eigenbasis of the SLD based on the exactly true value of $\theta_1$. But in reality, since we will carry out this measurement at the estimated value $\hat \theta_1 =\theta_1 +\theta_{1,\rm err}$, we must replace $\mathcal K$ by the FI for the SLD eigenbasis measurement evaluated at this value rather than the true $\theta_1$. We can express this FI as a Taylor expansion around the true value $\theta_1$ as
\be
\mathcal{J}_{\rm SLD}(\theta_1 +\theta_{1,\rm err}) = \mathcal{J}_{\rm SLD}(\theta_1) \,+\, \theta_{1,\rm err} ^2 \frac{\partial^2 \mathcal{J}_{\rm SLD}(\theta_1)}{\partial\theta_1^2} \,+\,\ldots
\ee
Here we do not have a first derivative term because $\mathcal{J}_{\rm SLD}$ has a maximum at $\theta_1$ equal to $\mathcal{K}$, so the first derivative must be zero. Moreover, since this is a maximum, the second derivative will be a negative constant with respect to $\theta_{1,\rm err}$. Therefore, we can rewrite the stage-two FI as
\be
\begin{split}
	\mathcal{J}_{\rm SLD}(\theta_1 +\theta_{1,\rm err}) = & \mathcal K\left(1 -O\left(\theta_{1, \rm err}^2\right)\right) \\
	= &\mathcal K \left(1-O\left(\frac{1}{N^\alpha \mathcal{J}_1}\right)\right)
\end{split}
\ee
where in the last step, we have used the fact that the mean squared error of the initial centroid estimate is approximately equal to the inverse of the FI for stage-one.
The total FI accumulated over stage-two is therefore $(N-N^\alpha) \mathcal K\left(1 -O\left(\frac{1}{N^\alpha J_1}\right)\right)$.
And since when $N$ is large
\be
\frac{1}{(N-N^\alpha) \mathcal K\left(1-O\left(\frac{1}{N^\alpha \mathcal{J}_1}\right)\right)}
\approx \frac{1}{N \mathcal{K}},
\ee
the variance approaches that of the optimal measurement. Based on the choice of the stage-one measurement, and its FI $\mathcal{J}_1$, one could optimize the choice of $\alpha$ such that the overall FI attained at the end of stage-two is maximized, for a given fixed $N$. A multi-stage adaptive quantum estimation algorithm is given in~\cite{Fujiwara_2006} where the result of each stage is used as input for the next one, which could lead to further improved performance in this non-asymptotic setting. However, finding the quantum optimal measurement for finite $N$ is being left open for future work. Note that in the context of this adaptive measurement, $N$ indicates the number of photons that are dedicated to estimating the centroid. If the ultimate goal is to estimate the separation, there also will be photons set aside for the estimation of the separation (e.g., using a SPADE aligned to the centroid estimate) \cite{Grace2020c}. %Now, of course, the estimate $\hat\theta_1$ obtained from stage one will not be totally efficient if the measurement in stage one is sub-optimal, so there should be some additional error proportional to the gap between the FI for stage one and the QFI, but this should be a sub-leading effect, and at leading order, the variance should approach the performance of the QFI attaining estimation scheme.

We can therefore apply this 2-step procedure for estimating the centroid if we can calculate the eigenvectors of the SLD. For 2 point sources of light, the non-zero entries of the SLD are given in Appendix
\ref{2pt-QFI-calculation}. For more than 2 light sources, including the case of a continuous line, we can obtain the SLD numerically, calculating it in the basis of HG-basis $1$-photon Fock states, as described in Appendix \ref{HG-basis-QFI-procedure}. It is important to note here that the SLDs not only depend on the centroid, but also on the separation (or, the object length, in case of a continuous line). Therefore, if we already know the separation, then we only need to estimate $\theta_1$ in stage one of our two-step adaptive scheme. But if we do not know the separation, then we also need to extract an initial estimate $\hat\theta_2$ for the separation from the measurement in the first stage. We can then switch to the second stage where we calculate the eigenbasis of the SLD in terms of $\hat\theta_1$ and $\hat\theta_2$ to get a good estimate for the centroid whose quality approaches the QFI for a large number of integrated photons.

%It is however worth mentioning that since direct imaging is not sub-optimal by a very large margin (of an order of magnitude or more), the actual performance of such a two or three stage adaptive scheme may not necessarily be better than raw direct imaging in actual practise, because a more elaborate scheme will also have additional noise.

\section{Conclusion}

We have carried out a detailed, systematic study of our ability to estimate the centroid of a linear array of incoherent light sources as well as a line-shaped object with uniform brightness. Our approach is easily extensible to estimating the centroid of more complex objects. We calculated the QFI for estimating the centroid and compared it with the FI for direct imaging as well as an image-plane HG SPADE. We described a two-stage readily-realizable measurement that would attain the QFI of centroid estimation.

Our key conclusions can be summarized as follows:

Direct imaging in the image plane is strictly sub-optimal compared to the QFI for centroid estimation, though the ratio of the QFI to the Direct imaging FI is less than an order of magnitude. Therefore, direct imaging should offer a fairly good estimate of the location of an object, as intuitively expected.

The gap between the performance of direct imaging and QFI is generally less for the continuum objects than for a constellation involving a small number of point-like emitters. This suggests that direct imaging performs closer to the optimal scheme for locating a full, uniformly bright object rather than one that has more internal features.

The performance of the HG SPADE is mostly worse than direct imaging for centroid estimation, except for some special limiting regimes where its FI approaches the direct imaging FI from below, or some small regions of parameter space where it marginally surpasses direct imaging. However, these regions where it slightly outperforms direct imaging are negligible, and the performance improvement is also too little to be of any practical significance.

We have also found that the HG mode performs very poorly when it is nearly aligned with the centroid. This means that if we employ a two-stage procedure for determining $\theta_1$ and $\theta_2$, where we first estimate $\theta_1$ and then use it to align the SPADE for determining $\theta_2$ in stage 2 as in~\cite{Grace2020c}, then we would not be able to extract much additional information about the centroid from the SPADE measurement in stage 2.

There is a complementarity between direct imaging and the HG SPADE. Direct imaging has a fairly good performance for centroid estimation, even though it is not the optimal measurement. But it performs very poorly for determining the separation in the sub-Rayleigh regime, and the Fisher information for the separation goes to zero when the separation approaches zero. The HG SPADE, on the other hand, is the optimal measurement for finding the separation, but it performs very poorly for determining the centroid, when it is nearly aligned with the centroid.

We have found an interesting scaling behavior for the QFI and direct imaging for the continuous line for large $\theta_2$, with a constant-factor gap. Specifically, we have found that the QFI and the direct imaging FI both scale as $1/\theta_2$ in this region. This makes very good intuitive sense: the larger the length of a continuous object, the greater the portion of the object that has spatially constant irradiance, and hence fewer information-bearing photons are available to estimate the centroid.

\section{Acknowledgements}
The authors thank Mankei Tsang and Ranjith Nair for valuable discussions. This work was supported by a Defense Advanced Research Projects Agency (DARPA) Defense Sciences Office (DSO) seedling project awarded under contract number W911NF2010039. QZ acknowledges the DARPA Young Faculty Award (YFA), Grant number N660012014029.

\appendix
\section{The diagonality of the FI and QFI for our centroid and separation parameters}
\label{FI-and-QFI-diagonality}

\subsection{The FI for Direct imaging}

First, consider direct imaging. For our choice of a Gaussian PSF, the density probability function $\Lambda(x)$ defined in (\ref{Lambda-def2})
is an even function around the centroid $x =\theta_1$.
It is a straightforward exercise to see that $\frac{\partial\Lambda(x)}{\partial\theta_1}$ is an odd function around the centroid, whereas $\frac{\partial\Lambda(x)}{\partial\theta_2}$ is even.
Their product is therefore an odd function around $\theta_1$, and the integral over $x$ from $-\infty$ to $\infty$ is therefore zero.

To see why $\partial\Lambda(x)/\partial\theta_1$ is odd and $\partial\Lambda(x)/\partial\theta_2$ is even, note that
if we have an even number of points, they come in pairs of the form
$x_s = \theta_1 \pm c_s\theta_2$,
in which $c_s$ is a factor that only depends on $s$.
If we have an odd number of points, then we have a point in the middle at $x=\theta_1$, and all the other points again come in such pairs with the same distance on either side of the centroid.
The partial derivative of $\Lambda(x)$ will therefore also contain pairs with contributions of the form
\be
\frac{\partial{|}\psi(x-\theta_1 \pm c_s\theta_2){|}^2}{\partial\theta_1}
\nn\ee\be
= -2{|}\psi(x-\theta_1 \pm c_s\theta_2){|}
\frac{\partial{|}\psi(x-\theta_1 \pm c_s\theta_2){|}}{\partial (x-\theta_1 \pm c_s\theta_2)}
\ee
Now, since ${|}\psi(x-\theta_1 \pm c_s\theta_2){|}$
is symmetric around $\theta_1 \mp c_s$,
$\frac{\partial{|}\psi(x-\theta_1 \pm c_s\theta_2){|}}{\partial (x-\theta_1 \pm c_s\theta_2)}$
is anti-symmetric.
And therefore, the sum over both elements of the pair with $\pm c_s\theta_2$ is an anti-symmetric function around $\theta_1$.
In case the number of points is odd, then the partial derivative of the middle point will also be an anti-symmetric function.
So $\partial\Lambda(x)/\partial\theta_1$ is odd around the centroid.

In contrast
\be
\frac{{|}\psi(x-\theta_1 \pm \theta_2){|}^2}{\partial\theta_2}
\nn\ee\be
= \pm 2c_s {|}\psi(x-\theta_1 \pm c_s\theta_2){|}
\frac{\partial{|}\psi(x-\theta_1 \pm c_s\theta_2){|}}{\partial (x-\theta_1 \pm c_s\theta_2)}
\ee
So now the two members of the pair have opposite signs. This, along with the fact that
$\frac{\partial{|}\psi(x-\theta_1 \pm c_s\theta_2){|}}{\partial (x-\theta_1 \pm c_s\theta_2)}$
is anti-symmetric around $\theta_1 \mp c_s\theta_2$,
means that the two members of each pair combine to give a symmetric function around $\theta_1$.
If the total number of points is odd, then there is also an additional point located in the middle at the centroid. But since its position does not depend on $\theta_2$, it does not contribute to $\partial\Lambda(x)/\partial\theta_2$.
Therefore, we conclude that $\partial\Lambda(x)/\partial\theta_2$ is an even function around $\theta_1$.

\subsection{The QFI}

For the QFI, the argument is somewhat similar, but now we need to think in terms of the density operator and its partial derivatives.
Recalling (\ref{rho1_definition}), the 1-photon part of the density operator is
\be
\rho_1
= \frac{1}{n} \sum_{s=1}^n {|}\psi_s\ra\la\psi_s{|}
=
\frac{1}{n}
\sum_{s=1}^n
{|}\psi(x-x_s)\ra\la\psi(x-x_s){|}
\ee
This again comes in pairs, and the density operator is symmetric in each pair.
Now, consider the partial derivative
\be
\frac{\partial\rho_1}{\partial\theta_1}
= \frac{1}{n}
\sum_{s=1}^n\Bigg(
\frac{ \partial{|}\psi(x-x_s)\ra}{\partial\theta_1}
\nn\ee\be
\la\psi(x-x_s){|}
+ {|}\psi(x-x_s)\ra \frac{\partial\la\psi(x-x_s){|}}{\partial\theta_1}\Bigg)
\ee
But ${|}\psi(x-x_s)\ra
= \int dx \, \psi(x-x_s) {|}x\ra$,
and therefore, it has the same even-odd parity as $\psi(x-x_s)$.
Likewise, the partial derivatives of this ket will have the same even-odd parity as the partial derivatives of the function $\psi(x-x_s)$.
Therefore, like $\partial\Lambda(x)/\partial\theta_1$ and $\partial\Lambda(x)/\partial\theta_2$, $\partial\rho_1/\partial\theta_1$ and $\partial\rho_1/\partial\theta_2$ will be odd and even, respectively. And therefore, the off-diagonal entries of the QFI evaluated in terms of the parameters $\theta_1$ and $\theta_2$ will be zero.

\subsection{The HG SPADE}

For the HG mode sorter, our physical set up is not symmetric unless the SPADE is perfectly aligned with the centroid.
Therefore, the above-mentioned symmetry arguments no longer hold, and the FI for a measurement in the HG basis is not diagonal in general. For the particular case of the SPADE being perfectly aligned with the centroid, we saw in section section \ref{SPADE-section}
that $\partial P(q)/\partial\theta_1$ becomes zero due to cancellations of the contributions from each member of the symmetric pair.
Therefore, the FI is not only diagonal, but all its entries other than the diagonal one corresponding to $\theta_2$ are zero.

However, when the SPADE is mis-aligned, we will not get a diagonal QFI matrix in general.
To see this, consider the specific example of the 2 point case. The probability function is
\be
P(q)
= \frac{1}{2q!}
\left(\exp\left(-Q_1\right) \, Q_1^q
\,+\, \exp\left(-Q_2\right)\, Q_2^q\right)
\ee
where $Q_1 = \frac{(\theta_1 +\theta_2/2)^2}{4\sigma^2}$
and $Q_2 =\frac{(\theta_1 -\theta_2/2)^2}{4\sigma^2}$.
The two terms are clearly not symmetric or anti-symmetric, since they have different Gaussian decay factors as well as $(\theta_1 \pm \theta_2)^q$ which will be different for both points.

\section{The single point case}
\label{single-point-case-appendix}

\subsection{The FI for direct imaging}

For a single point, the probability function $\Lambda(x)$ defined in
(\ref{Lambda-def2}) becomes
\be
\Lambda_1(x) = {|}\psi(x -\theta_1){|}^2
\ee
The Fisher information for direct imaging is then
\begin{equation}
	\begin{split}
		\mathcal{J}_{\rm 1-pt}
		=& \int dx\, \frac{N}{\Lambda(x)}
		\left(\frac{\partial\Lambda(x)}{\partial\theta_1}\right)^2
		\\
		=& 4N
		\int dx\,
		\left(\frac{\partial{|}\psi(x-\theta_1){|}}{\partial\theta_1}\right)^2
		\\
		=&4N\Delta k^2
		\label{direct-imaging-FI-single-point-general}
	\end{split}
\end{equation}

For our Gaussian PSF defined in (\ref{PSF-definition}), the result is
\be
\mathcal{J}_{\rm 1-pt} = \frac{N}{\sigma^2}
\ee

\subsection{The QFI}

When we only have one point, the single-photon part of the density matrix is simply one-dimensional
\be
\rho_1 = {|}\psi(x-\theta_1)\ra\la\psi(x-\theta_1){|}
\ee
The partial derivative of this is
\be
\frac{\partial\rho_1}{\partial\theta_1}
= \Bigg(\frac{\partial{|}\psi(x-\theta_1)\ra}{\partial\theta_1}
\nn\ee\be
\la\psi(x-\theta_1){|}
+ {|}\psi(x-\theta_1)\ra
\frac{\partial\la\psi(x-\theta_1){|}}{\partial\theta_1}\Bigg)
\nn\ee\be
= -{|}\psi'(x-\theta_1)\ra\la \psi(x-\theta_1){|}
-{|}\psi(x-\theta_1)\ra\la \psi'(x-\theta_1){|}
\label{rho-single-point-single-point}
\ee
since $\partial\psi(x-\theta_1)/\partial\theta_1
= -\psi'(x-\theta_1)$
where $\psi'$ is the derivative of $\psi$.
It is straightforward to see that ${|}\psi(x-\theta_1)\ra$
and ${|}\psi'(x-\theta_1)\ra$
are orthogonal states for any symmetric $\psi(x)$:
\be
\la\psi(x-\theta_1){|}\psi'(x-\theta_1)\ra
= \int dx\, \psi(x-\theta_1) \, \frac{\partial\psi(x-\theta_1)}{\partial\theta_1}
\nn\ee\be
= 0
\ee
We thus have an orthogonal basis and only need to normalize $\partial{|}\psi(x-\theta_1)\ra/\partial\theta_1$.
Our orthonormal basis is thus
\be
{|}e_1\ra \equiv |\psi(x-\theta_1)\ra
\ee\be
{|}e_2\ra \equiv \frac{1}{\Delta k} {|}\psi'(x-\theta_1)\ra
\ee
where $\Delta k=\sqrt{\int dx |\psi'(x-\theta_1)|^2}$ is a normalization factor and is equal to the square root of $\Delta k^2$, defined in (\ref{Delta-k2-def}). The density operator in this eigen basis is simply $\rho_1 = {|}e_1\ra\la e_1{|}$ with eigen values $D_1 = 1$ and $D_2 = 0$ corresponding to ${|}e_1\ra$ and ${|}e_2\ra$.
Recalling (\ref{sld}) and (\ref{rho-single-point-single-point}),
the symmetric logarithmic derivative is then
\be
\sL
= 2\Delta k
\left( {|}e_2\ra\la e_1{|}
+ {|}e_1\ra\la e_2{|}\right)
\ee
The QFI is then
\be
\mathcal{K}_{\rm 1-pt}
= N tr\left(\rho \sL^2\right)
= 4N\Delta k^2
\ee
We see that this is equal to the FI for direct imaging in
(\ref{direct-imaging-FI-single-point-general}) for any PSF.

\subsection{The HG SPADE FI}
\label{HG-FI-single-point}

The probability function for the $q$th HG mode for light coming from a single point with a Gaussian PSF is given by
\be
P_q =
\exp\left(-Q\right) \, \frac{Q^q}{q!}
\ee
where $Q = \frac{\theta_1 ^2}{4\sigma^2}$.
The FI for the $q$th mode is then
\be
\mathcal{J}_{q, {\rm HG, 1-pt}}
= \frac{N}{P_q} \left(\frac{\partial P_q}{\partial\theta_1}\right)^2
\ee
It is a straightforward exercise to calculate this and carry out the sum over the whole series in $q$, and the result is
\be
\mathcal{J}_{\rm HG, 1-pt}
= \frac{N}{\sigma^2}
\ee
which is equal to the QFI as well as the FI for direct imaging.

\section{The QFI for 2 points}
\label{2pt-QFI-calculation}

This calculation has been explained by Tsang {\em et al.} in their paper. Therefore, we will only summarize their method while clarifying one or two points.

From (\ref{rho1_definition}), the single-photon part of the density matrix for the 2 point case is
\be
\rho_1 = \frac{1}{2}
\left({|}\psi_1\ra\la\psi_1{|}
+ {|}\psi_2\ra\la\psi_2{|}\right)
\label{rho-2pt-def}
\ee
However, for the QFI, we need to work in an orthonormal eigenbasis that spans the whole space spanned by ${|}\psi_1\ra$ and ${|}\psi_2\ra$ as well as their partial derivatives with respect to $\theta_1$.
And it turns out that while ${|}\psi_1\ra$ and ${|}\psi_2\ra$ individually have norm 1, they are not mutually orthogonal in general
\be
\delta \equiv \la\psi_1{|}\psi_2\ra
=\la\psi_2{|}\psi_1\ra
\neq 0
\label{delta-definition}
\ee
for a real valued $\psi(x)$.
Therefore, we first need to express $\rho_1$ in an orthonormal basis
\be
\rho_1 = D_1 {|}e_1\ra\la e_1{|}
+ D_2{|} e_2\ra\la e_2{|}
\ee
To find the eigenvalues $D_i$ and the eigenstates ${|}\psi_i\ra$,
we write down a $2\times 2$ matrix of the inner products $\la\psi_i{|}\rho{|}\psi_j\ra$
\be
\begin{pmatrix}
	1 & \delta \\
	\delta & 1
\end{pmatrix}
\label{dot-product-matrix}
\ee
The normalized eigenvectors of this matrix give us
an orthogonal set of functions, and the square roots of the eigenvalues give us the normalization factors.
We find that the eigenvalues are $1\pm \delta$,
with the eigenvectors $\frac{1}{\sqrt{2}}(1, \pm 1)$.
Therefore, our orthonormal basis of states spanning ${|}\psi_1\ra$ and ${|}\psi_2\ra$ is
\be
{|}e_1\ra = \frac{1}{\sqrt{2(1-\delta)}} ({|}\psi_1\ra -{|}\psi_2\ra)
\ee\be
{|}e_2\ra = \frac{1}{\sqrt{2(1+\delta)}} ({|}\psi_1\ra +{|}\psi_2\ra)
\ee
It is a straightforward exercise to see that these are also the eigenstates of our density operator $\rho_1$,
and that the eigenvalues $D_i$ of $\rho_1$
are the corresponding eigenvalues of the matrix of inner products (\ref{dot-product-matrix}) divided by 2:
\be
D_1 = \frac{1-\delta}{2}
\ee\be
D_2 = \frac{1+\delta}{2}
\ee
This division by 2 is simply the 1/2 factor in front of ${|}\psi_1\ra\la\psi_1{|} +{|}\psi_2\ra\la\psi_2{|}$.

However, $\frac{\partial\rho_1}{\partial\theta_1}$
also contains the derivatives of ${|}\psi_1\ra$ and ${|}\psi_2\ra$.
Therefore, we need to extend our eigenbasis to span these states too. We therefore include the derivatives of ${|}\psi_i\ra$ and carry out an orthogonalization procedure. This gives us the following additional states
\be
{|}e_3\ra = \frac{1}{c_3}
\Bk{\frac{\Delta k}{\sqrt{2}}\bk{{|}\psi_{11}\ra + {|}\psi_{22}\ra}
	- \frac{\gamma}{\sqrt{1-\delta}}{|} e_1\ra},
\ee\be
{|} e_4\ra = \frac{1}{c_4}
\Bk{\frac{\Delta k}{\sqrt{2}}\bk{{|} \psi_{11}\ra - {|} \psi_{22}\ra} +
	\frac{\gamma}{\sqrt{1+\delta}}{|} e_2\ra},
\ee
where $\Delta k^2$ and $\gamma$ were defined in (\ref{Delta-k2-def}) and (\ref{gamma-def}), which are reproduced here for the reader's convenience
\be
\Delta k^2 = \int_{-\infty}^\infty dx \left[\frac{\partial \psi(x)}{\partial x}\right]^2
\ee
and
\be
\gamma = \int_{-\infty}^\infty dx \frac{\partial \psi(x)}{\partial x} \psi(x -\theta_2)
\ee
The other quantities defined here are
\be
{|}\psi_{11}\ra \equiv \frac{1}{\Delta k}\int dx \,
\frac{\partial \psi(x-x_1)}{\partial x_1} {|} x\ra,
\ee\be
{|}\psi_{22}\ra \equiv \frac{1}{\Delta k}\int dx \, \frac{\partial \psi(x-x_2)}{\partial x_2}{|} x\ra,
\ee\be
c_3 \equiv \bk{\Delta k^2 + b^2 - \frac{\gamma^2}{1-\delta}}^{1/2},
\ee\be
c_4 \equiv \bk{\Delta k^2 - b^2 - \frac{\gamma^2}{1+\delta}}^{1/2},
\ee\be
b^2 \equiv \int dx\, \frac{\partial \psi(x-x_1)}{\partial x_1} \frac{\partial \psi(x-x_2)}{\partial x_2},
\ee
and $\delta$ was defined in (\ref{delta-definition}).

Since $\rho_1 = D_1{|}e_1\ra\la e_1{|} + D_2{|}e_2\ra\la e_2{|}$,
we get
\be
D_3 = D_4= 0
\ee
Having found all the eigenbasis states and the eigenvalues of $\rho_1$, it is now a simple exercise to use the formula (\ref{sld}) for the SLD and compute the QFI, with the results given in (\ref{K11-2pt-result-apxC}) and (\ref{K22-2pt-result-apxC}).
For reference, the non-zero entries of the SLD with respect to the centroid in the ${|}e_i\rangle$ ($i = 1\ldots 4$) basis. The SLD with respect to the centroid has the non-zero entries are as follows:
\be
\sL_{1, 12}
= \frac{2\gamma\delta}{\sqrt{1-\delta^2}}
\ee\be
\sL_{1, 14}
= \frac{2c_4}{\sqrt{1-\delta}}
\ee\be
\sL_{1, 23}
=\frac{2c_3}{\sqrt{1+\delta}}
\ee
The non-zero entries of the SLD with respect to the separation are
\be
\sL_{2, 11}
= -\frac{\gamma}{1-\delta}
\ee\be
\sL_{2, 13}
= -\frac{c_3}{\sqrt{1-\delta}}
\ee\be
\sL_{2, 22}
= \frac{\gamma}{1+\delta}
\ee\be
\sL_{2, 24}
= -\frac{c_4}{\sqrt{1+\delta}}
\ee
From these, it is a straightforward exercise to calculate the QFI. For the centroid, we obtain
\be
\mathcal{K}_{11} = 4N(\Delta k^2 -\gamma^2)
\label{K11-2pt-result-apxC}
\ee
and for the separation,
\be
k_{22} = N\Delta k^2
\label{K22-2pt-result-apxC}
\ee

%I think this appendix is not finished

\section{Calculating the QFI for more than 2 points}
\label{HG-basis-QFI-procedure}

Calculating the QFI analytically for more than 2 points by employing Tsang {\em et al's} procedure for 2 points becomes a very complicated process, since it requires diagonalizing larger and larger matrices as the number of points is increased. Even doing this numerically is a very involved process, and in fact we soon start running into floating point errors
when we go to about 10 or so points. It also does not allow us to calculate the QFI for a continuous line. We therefore follow the more efficient numerical approach employed in~\cite{Zachary2019}. The idea is that instead of working with states ${|}\psi_i\ra$ and their derivatives and carrying out a laborious diagonalization process, we work in the HG basis. We only need to consider the first few HG modes, since higher order modes have diminishing contributions. The HG SPADE calculations in this paper have been carried out with 50 HG modes, and we have checked that this is more than enough for the results to converge for the parameters being considered.

Specifically, we express our states $\psi(x-x_s)$ in the HG basis, which gives
\be
\psi(x-x_s)
= \sum_{q=0}^\infty \exp\left(-\frac{x_s ^2}{8\sigma^2}\right)
\frac{x_s ^q}{\sqrt{q!}}
\, \phi_q(x)
\ee
where $\phi_q(x)$ are the HG functions.
We then express our density matrix $\rho_1$ and its derivative $\partial\rho_1 /\partial\theta_1$ in this basis.
Since $\phi_q(x)$ does not depend on the location of the individual points or the centroid, the partial derivatives do not change the basis, and therefore, we do not have to carry out any orthogonalization procedure to find our additional basis states. We simply numerically calculate the eigenvectors and eigenvalues for the $\rho_1$
in the basis of the first 50 HG modes (or whatever other number we decide to consider). We then use the formula
(\ref{sld}) to obtain the SLD, and calculate the QFI from
$N \trace\left(\rho \sL^2\right)$.

\section{The QFI and the direct imaging FI for the continuum case and its scaling behavior}
\label{scaling-appendix}

Here we show why the QFI and direct imaging FI for the centroid scale as $1/\theta_2$ for large $\theta_2$ in the continuum case of an infinite number of emitters. First, consider direct imaging and recall the definition of the probability function $\Lambda(x)$ in
(\ref{Lambda-def2}). This is an average over $m$ points, and for a continuous line, we replace it by an integral
\be
\Lambda_{\rm line}(x) = \frac{\int_{y=-\theta_2/2}^{\theta_2/2}
	{|}\psi(x-\theta_1 -y){|}^2 dy}
{\theta_2}
\label{Lambda-line}
\ee
For $\theta_2$ sufficiently large compared to $\sigma$,
this, being an integral of a sharply peaked but smooth function, should give a nearly flat function that is constant over the length of the line and zero elsewhere, but with smooth edges with width of order $\sigma$. For our Gaussian PSF, we get
\be
\Lambda_{\rm line}(x)= \frac{1}{2\theta_2}
\left(
{\rm erf}\Bigg(
\frac{x -\theta_1 +\theta_2/2}
{\sqrt{2} \sigma}\right)
\nn\ee\be
-{\rm erf}\left(
\frac{x-\theta_1 -\theta_2/2}
{\sqrt{2} \sigma}\Bigg)
\right)
\label{Lambda-line2}
\ee
which is indeed nearly $1/\theta_2$ over the line and almost zero elsewhere, but has smoothly falling edges.
The other ingredient we need for the FI is the partial derivative of this with respect to $\theta_1$, which can easily be computed using the fundamental theorem of calculus. Since
$\frac{\partial {|}\psi(x-\theta_1 -y){|}^2}{\partial \theta_1}
= \frac{\partial {|}\psi(x-\theta_1 -y){|}^2}{\partial y}$,
we obtain $\frac{\partial\Lambda_{\rm line}(x)}{\partial\theta_1}$ by removing the integral and evaluating ${|}\psi(x-\theta_1 -y){|}^2$ at the $y$ values of the end points:
\be
\frac{\partial\Lambda_{\rm line}(x)}{\partial\theta_1}
= \frac{1}{\theta_2}
\big({|}\psi(x-\theta_1 -\theta_2/2){|}^2
\nn\ee\be
-{|}\psi(x-\theta_1 +\theta_2/2){|}^2\big)
\label{Lambda-derivative-line}
\ee
If $\theta_2$ is sufficiently large compared to $\sigma$,
this is the difference between two non-overlapping sharply peaked but smooth functions at the edges of the line.
The square of this will therefore be a sum of two even more sharply peaked functions at the edges of the line.
The integral over $x$ to evaluate FI therefore only gets noticeable contributions near the edges of the line, and hence the distance between the edges does not have any bearing on it. This leaves the $1/\theta_2$ factor in front as the main $\theta_2$ dependent part in (\ref{Lambda-derivative-line}).
This, along with the $1/\theta_2$ scaling of $\Lambda(x)$ in (\ref{Lambda-line2}), means that the FI should scale as $1/\theta_2$. It is worth noting that this reasoning should hold equally for any other PSF that has a high peak in the center and quickly but smoothly falls to zero away from it.

A somewhat similar argument can be made for the scaling of the QFI in the large $\theta_2$ region. The sum in (\ref{rho1_definition}) for the density matrix gets replaced by an integral over the line. Recalling the definition of ${|}\psi_s\ra$ in (\ref{psi_s-ket-definition}),
this integral is
\be
\begin{split}
\rho_1 =&
\frac{1}{\theta_2}
\int dx\, dx' \, \int_{y=-\theta_2/2}^{\theta_2/2} dy \, \psi(x-\theta_1-y)\\
&\times \psi(x'-\theta_1-y) {|}x\ra\la x'{|}
\end{split}
\label{rho1-definition-line}
\ee
The partial derivative of this with respect to $\theta_1$ can again be obtained from the fundamental theorem of calculus by removing the integral over $y$ and evaluating this at the end points:
\be
\rho_1 =
\int dx\, dx' \,\frac{1}{\theta_2}\Big[
\psi(x-\theta_1 -\theta_2/2) \, \psi(x'-\theta_1-\theta_2/2)
\nn\ee\be
- \psi(x-\theta_1+\theta_2/2) \, \psi(x'-\theta_1+\theta_2/2)\Big]
{|}x\ra\la x'{|}
\label{rho1-derivative-definition-line}
\ee
For $\theta_2$ sufficiently larger than $\sigma$, the two terms will sharply peak when both $x$ and $x'$ are simultaneously equal to the end point locations of the line at $\theta_1 +\theta_2/2$
and $\theta_1 -\theta_2/2$. Elsewhere they will be nearly zero.
The height and width of these peaks will not depend on $\theta_2$,
and therefore the scaling of $\partial\rho/\partial\theta_1$
in terms of $\theta_2$ arises almost entirely from the $1/\theta_2$ factor in front.

As for the scaling of $\rho$, we can return to (\ref{rho1-definition-line}).
Again, focusing on $\theta_2$ sufficiently larger than $\sigma$,
we note that with $\psi(x-y)$ and $\psi(x'-y)$ sharply peaked at $y=x$ and $y=x'$, performing the integral over $y$
will give us a function that sharply peaks at $x=x'$ provided $x$ and $x'$ lie somewhere on our line between $-\theta_2/2$ and $\theta_2/2$. For our Gaussian point spread function, we get
\be
\rho_1 =
\frac{1}{\theta_2}
\int dx\, dx' \,
\frac{\exp\left(-\frac{(x-x')^2}{8\sigma^2}\right)
}{2}
\nn\ee\be
\left(
{\rm erf}\left(\frac{\theta+x+x')}{2^{3/2}\sigma}\right)
-{\rm erf}\left(\frac{-\theta+x+x'}
{2^{3/2}\sigma}\right)
\right)
\ee
in which we indeed have a sharply peaked Gaussian involving $(x-x')$, and the erf functions are simply step functions with smooth edges, and hence do not contribute to the scaling in terms of $\theta_2$. This leaves the $1/\theta_2$ in front as the only factor that contributes to the scaling.
Since the QFI is given as
$K = tr(\rho\sL ^2)$
and
satisfies the relation (\ref{SLD-relation})
it is clear that the overall scaling is essentially of two powers of $\partial\rho/\partial\theta_1$ and an inverse power of $\rho$, so overall we get $1/\theta_2$. Like the argument for the direct imaging FI, this reasoning for the scaling of the QFI should also hold for any other smooth PSF that has a sufficiently sharp peak.

\section{HG SPADE FI for the continuum case}
\label{HG-SPADE-for-line}

For the continuum case, we replace the sum in (\ref{P-def}) by an integral.
Recalling (\ref{P_s-def}),
we get
\be
P_l(q) = \frac{1}{\theta_2} \int_{y =-\theta_2/2}^{\theta_2/2}
dy\,\exp\left(-Q(y)\right) \frac{Q(y) ^q}{q!}
\label{P_l-def}
\ee
where
\be
Q(y)
= \frac{\theta_1 +y}{4\sigma^2}
\ee
And the partial derivative with respect to $\theta_1$ can be obtained, according to the fundamental theorem of calculus, by just removing the integral over $y$ and evaluating at the end points:
\be
\frac{\partial P_l(q)}{\partial\theta_1} = \frac{1}{\theta_2}
\left(\exp\left(-Q_+\right) \frac{Q_+ ^q}{q!}
-\exp\left(-Q_-\right) \frac{Q_- ^q}{q!} \right)
\label{P_l-derivative-def}
\ee
where
\be
Q_\pm = \frac{(\theta_1 \pm \theta_2/2)^2}{4\sigma^2}
\ee
It is now straightforward to write down the FI contribution for each HG mode, and summing over all the modes gives us the total FI
\be
\mathcal{J}_{{\rm HG}, l, 11}
= N\sum_{q=0}^\infty \frac{\left(\exp\left(-Q_+\right) \frac{Q_+ ^q}{q!}
	-\exp\left(-Q_-\right) \frac{Q_- ^q}{q!} \right)^2}
{\theta_2 \int_{y =-\theta_2/2}^{\theta_2/2}
	dy\,\exp\left(-Q(y)\right) \frac{Q(y) ^q}{q!}}
\label{J_hg_line-total}
\ee

\bibliography{ref_centroid}

\begin{thebibliography}{33}%
\makeatletter
\providecommand \@ifxundefined [1]{%
 \@ifx{#1\undefined}
}%
\providecommand \@ifnum [1]{%
 \ifnum #1\expandafter \@firstoftwo
 \else \expandafter \@secondoftwo
 \fi
}%
\providecommand \@ifx [1]{%
 \ifx #1\expandafter \@firstoftwo
 \else \expandafter \@secondoftwo
 \fi
}%
\providecommand \natexlab [1]{#1}%
\providecommand \enquote  [1]{``#1''}%
\providecommand \bibnamefont  [1]{#1}%
\providecommand \bibfnamefont [1]{#1}%
\providecommand \citenamefont [1]{#1}%
\providecommand \href@noop [0]{\@secondoftwo}%
\providecommand \href [0]{\begingroup \@sanitize@url \@href}%
\providecommand \@href[1]{\@@startlink{#1}\@@href}%
\providecommand \@@href[1]{\endgroup#1\@@endlink}%
\providecommand \@sanitize@url [0]{\catcode `\\12\catcode `\$12\catcode
  `\&12\catcode `\#12\catcode `\^12\catcode `\_12\catcode `\%12\relax}%
\providecommand \@@startlink[1]{}%
\providecommand \@@endlink[0]{}%
\providecommand \url  [0]{\begingroup\@sanitize@url \@url }%
\providecommand \@url [1]{\endgroup\@href {#1}{\urlprefix }}%
\providecommand \urlprefix  [0]{URL }%
\providecommand \Eprint [0]{\href }%
\providecommand \doibase [0]{https://doi.org/}%
\providecommand \selectlanguage [0]{\@gobble}%
\providecommand \bibinfo  [0]{\@secondoftwo}%
\providecommand \bibfield  [0]{\@secondoftwo}%
\providecommand \translation [1]{[#1]}%
\providecommand \BibitemOpen [0]{}%
\providecommand \bibitemStop [0]{}%
\providecommand \bibitemNoStop [0]{.\EOS\space}%
\providecommand \EOS [0]{\spacefactor3000\relax}%
\providecommand \BibitemShut  [1]{\csname bibitem#1\endcsname}%
\let\auto@bib@innerbib\@empty
%</preamble>
\bibitem [{\citenamefont {Lord~Rayleigh}(1879)}]{LordRayleigh1879}%
  \BibitemOpen
  \bibfield  {author} {\bibinfo {author} {\bibfnamefont {F.~R.~S.}\
  \bibnamefont {Lord~Rayleigh}},\ }\bibfield  {title} {\bibinfo {title} {Xxxi.
  investigations in optics, with special reference to the spectroscope},\
  }\href {https://doi.org/10.1080/14786447908639684} {\bibfield  {journal}
  {\bibinfo  {journal} {Philosophical Magazine Series 5}\ }\textbf {\bibinfo
  {volume} {8}},\ \bibinfo {pages} {261} (\bibinfo {year} {1879})}\BibitemShut
  {NoStop}%
\bibitem [{\citenamefont {Van~Trees}\ \emph {et~al.}(2013)\citenamefont
  {Van~Trees}, \citenamefont {Bell},\ and\ \citenamefont
  {Tian}}]{VanTrees2013}%
  \BibitemOpen
  \bibfield  {author} {\bibinfo {author} {\bibfnamefont {H.~L.}\ \bibnamefont
  {Van~Trees}}, \bibinfo {author} {\bibfnamefont {K.~L.}\ \bibnamefont
  {Bell}},\ and\ \bibinfo {author} {\bibfnamefont {Z.}~\bibnamefont {Tian}},\
  }\href@noop {} {\emph {\bibinfo {title} {Detection, Estimation, and
  Modulation Theory, Part I}}},\ \bibinfo {edition} {2nd}\ ed.\ (\bibinfo
  {publisher} {Wiley},\ \bibinfo {year} {2013})\BibitemShut {NoStop}%
\bibitem [{\citenamefont {Mari}\ \emph {et~al.}(2012)\citenamefont {Mari},
  \citenamefont {Tamburini}, \citenamefont {Swartzlander}, \citenamefont
  {Bianchini}, \citenamefont {Barbieri}, \citenamefont {Romanato},\ and\
  \citenamefont {Thid\'{e}}}]{Mari2012-ps}%
  \BibitemOpen
  \bibfield  {author} {\bibinfo {author} {\bibfnamefont {E.}~\bibnamefont
  {Mari}}, \bibinfo {author} {\bibfnamefont {F.}~\bibnamefont {Tamburini}},
  \bibinfo {author} {\bibfnamefont {G.~A.}\ \bibnamefont {Swartzlander}},
  \bibinfo {author} {\bibfnamefont {A.}~\bibnamefont {Bianchini}}, \bibinfo
  {author} {\bibfnamefont {C.}~\bibnamefont {Barbieri}}, \bibinfo {author}
  {\bibfnamefont {F.}~\bibnamefont {Romanato}},\ and\ \bibinfo {author}
  {\bibfnamefont {B.}~\bibnamefont {Thid\'{e}}},\ }\bibfield  {title} {\bibinfo
  {title} {Sub-rayleigh optical vortex coronagraphy},\ }\href
  {https://doi.org/10.1364/OE.20.002445} {\bibfield  {journal} {\bibinfo
  {journal} {Opt. Express}\ }\textbf {\bibinfo {volume} {20}},\ \bibinfo
  {pages} {2445} (\bibinfo {year} {2012})}\BibitemShut {NoStop}%
\bibitem [{\citenamefont {Xu}\ \emph {et~al.}(2018)\citenamefont {Xu},
  \citenamefont {Wang},\ and\ \citenamefont {He}}]{Xu2018-li}%
  \BibitemOpen
  \bibfield  {author} {\bibinfo {author} {\bibfnamefont {B.}~\bibnamefont
  {Xu}}, \bibinfo {author} {\bibfnamefont {Z.}~\bibnamefont {Wang}},\ and\
  \bibinfo {author} {\bibfnamefont {J.}~\bibnamefont {He}},\ }\bibfield
  {title} {\bibinfo {title} {Super-resolution imaging via aperture modulation
  and intensity extrapolation},\ }\href@noop {} {\bibfield  {journal} {\bibinfo
   {journal} {Scientific reports}\ }\textbf {\bibinfo {volume} {8}},\ \bibinfo
  {pages} {1} (\bibinfo {year} {2018})}\BibitemShut {NoStop}%
\bibitem [{\citenamefont {Rust}\ \emph {et~al.}(2006)\citenamefont {Rust},
  \citenamefont {Bates},\ and\ \citenamefont {Zhuang}}]{Rust2006-mf}%
  \BibitemOpen
  \bibfield  {author} {\bibinfo {author} {\bibfnamefont {M.~J.}\ \bibnamefont
  {Rust}}, \bibinfo {author} {\bibfnamefont {M.}~\bibnamefont {Bates}},\ and\
  \bibinfo {author} {\bibfnamefont {X.~W.}\ \bibnamefont {Zhuang}},\ }\bibfield
   {title} {\bibinfo {title} {{Sub-diffraction-limit imaging by stochastic
  optical reconstruction microscopy (STORM)}},\ }\href {https://doi.org/Doi
  10.1038/Nmeth929} {\bibfield  {journal} {\bibinfo  {journal} {Nature
  Methods}\ }\textbf {\bibinfo {volume} {3}},\ \bibinfo {pages} {793} (\bibinfo
  {year} {2006})}\BibitemShut {NoStop}%
\bibitem [{\citenamefont {Vicidomini}\ \emph {et~al.}(2018)\citenamefont
  {Vicidomini}, \citenamefont {Bianchini},\ and\ \citenamefont
  {Diaspro}}]{Vicidomini2018-tm}%
  \BibitemOpen
  \bibfield  {author} {\bibinfo {author} {\bibfnamefont {G.}~\bibnamefont
  {Vicidomini}}, \bibinfo {author} {\bibfnamefont {P.}~\bibnamefont
  {Bianchini}},\ and\ \bibinfo {author} {\bibfnamefont {A.}~\bibnamefont
  {Diaspro}},\ }\bibfield  {title} {\bibinfo {title} {Sted super-resolved
  microscopy},\ }\href@noop {} {\bibfield  {journal} {\bibinfo  {journal}
  {Nature methods}\ }\textbf {\bibinfo {volume} {15}},\ \bibinfo {pages} {173}
  (\bibinfo {year} {2018})}\BibitemShut {NoStop}%
\bibitem [{\citenamefont {Helstrom}(1976)}]{Helstrom1976}%
  \BibitemOpen
  \bibfield  {author} {\bibinfo {author} {\bibfnamefont {C.~W.}\ \bibnamefont
  {Helstrom}},\ }\href@noop {} {\emph {\bibinfo {title} {Quantum Detection and
  Estimation Theory}}}\ (\bibinfo  {publisher} {Academic Press, New York},\
  \bibinfo {year} {1976})\BibitemShut {NoStop}%
\bibitem [{\citenamefont {Tsang}\ \emph {et~al.}(2016)\citenamefont {Tsang},
  \citenamefont {Nair},\ and\ \citenamefont {Lu}}]{Tsang2016b}%
  \BibitemOpen
  \bibfield  {author} {\bibinfo {author} {\bibfnamefont {M.}~\bibnamefont
  {Tsang}}, \bibinfo {author} {\bibfnamefont {R.}~\bibnamefont {Nair}},\ and\
  \bibinfo {author} {\bibfnamefont {X.-M.}\ \bibnamefont {Lu}},\ }\bibfield
  {title} {\bibinfo {title} {Quantum theory of superresolution for two
  incoherent optical point sources},\ }\href
  {https://doi.org/10.1103/PhysRevX.6.031033} {\bibfield  {journal} {\bibinfo
  {journal} {Phys. Rev. X}\ }\textbf {\bibinfo {volume} {6}},\ \bibinfo {pages}
  {031033} (\bibinfo {year} {2016})}\BibitemShut {NoStop}%
\bibitem [{\citenamefont {Kerviche}\ \emph {et~al.}(2017)\citenamefont
  {Kerviche}, \citenamefont {Guha},\ and\ \citenamefont {Ashok}}]{KGA17}%
  \BibitemOpen
  \bibfield  {author} {\bibinfo {author} {\bibfnamefont {R.}~\bibnamefont
  {Kerviche}}, \bibinfo {author} {\bibfnamefont {S.}~\bibnamefont {Guha}},\
  and\ \bibinfo {author} {\bibfnamefont {A.}~\bibnamefont {Ashok}},\ }\bibfield
   {title} {\bibinfo {title} {Fundamental limit of resolving two point sources
  limited by an arbitrary point spread function},\ }in\ \href@noop {} {\emph
  {\bibinfo {booktitle} {2017 IEEE International Symposium on Information
  Theory (ISIT)}}}\ (\bibinfo {organization} {IEEE},\ \bibinfo {year} {2017})\
  pp.\ \bibinfo {pages} {441--445}\BibitemShut {NoStop}%
\bibitem [{\citenamefont {J.~{\u R}eh{\'a}{\u c}ek}\ and\ \citenamefont
  {S{\'a}nchez-Soto}(2017)}]{Rehacek2017a}%
  \BibitemOpen
  \bibfield  {author} {\bibinfo {author} {\bibfnamefont {B.~S. Z.~H.}\
  \bibnamefont {J.~{\u R}eh{\'a}{\u c}ek}, \bibfnamefont {M.~Pa{\'u}r}}\ and\
  \bibinfo {author} {\bibfnamefont {L.~L.}\ \bibnamefont {S{\'a}nchez-Soto}},\
  }\bibfield  {title} {\bibinfo {title} {Optimal measurements for resolution
  beyond the rayleigh limit},\ }\href@noop {} {\bibfield  {journal} {\bibinfo
  {journal} {Opt. Lett. 42}\ ,\ \bibinfo {pages} {231}} (\bibinfo {year}
  {2017})}\BibitemShut {NoStop}%
\bibitem [{\citenamefont {Dutton}\ \emph {et~al.}(2019)\citenamefont {Dutton},
  \citenamefont {Kerviche}, \citenamefont {Ashok},\ and\ \citenamefont
  {Guha}}]{Zachary2019}%
  \BibitemOpen
  \bibfield  {author} {\bibinfo {author} {\bibfnamefont {Z.}~\bibnamefont
  {Dutton}}, \bibinfo {author} {\bibfnamefont {R.}~\bibnamefont {Kerviche}},
  \bibinfo {author} {\bibfnamefont {A.}~\bibnamefont {Ashok}},\ and\ \bibinfo
  {author} {\bibfnamefont {S.}~\bibnamefont {Guha}},\ }\bibfield  {title}
  {\bibinfo {title} {Attaining the quantum limit of superresolution in imaging
  an object's length via predetection spatial-mode sorting},\ }\href
  {https://doi.org/10.1103/PhysRevA.99.033847} {\bibfield  {journal} {\bibinfo
  {journal} {Phys. Rev. A}\ }\textbf {\bibinfo {volume} {99}},\ \bibinfo
  {pages} {033847} (\bibinfo {year} {2019})}\BibitemShut {NoStop}%
\bibitem [{\citenamefont {Ang}\ \emph {et~al.}(2017)\citenamefont {Ang},
  \citenamefont {Nair},\ and\ \citenamefont {Tsang}}]{Ang2016}%
  \BibitemOpen
  \bibfield  {author} {\bibinfo {author} {\bibfnamefont {S.~Z.}\ \bibnamefont
  {Ang}}, \bibinfo {author} {\bibfnamefont {R.}~\bibnamefont {Nair}},\ and\
  \bibinfo {author} {\bibfnamefont {M.}~\bibnamefont {Tsang}},\ }\bibfield
  {title} {\bibinfo {title} {Quantum limit for two-dimensional resolution of
  two incoherent optical point sources},\ }\href
  {https://doi.org/10.1103/PhysRevA.95.063847} {\bibfield  {journal} {\bibinfo
  {journal} {Phys. Rev. A}\ }\textbf {\bibinfo {volume} {95}},\ \bibinfo
  {pages} {063847} (\bibinfo {year} {2017})}\BibitemShut {NoStop}%
\bibitem [{\citenamefont {Yu}\ and\ \citenamefont {Prasad}(2018)}]{Prasad2018}%
  \BibitemOpen
  \bibfield  {author} {\bibinfo {author} {\bibfnamefont {Z.}~\bibnamefont
  {Yu}}\ and\ \bibinfo {author} {\bibfnamefont {S.}~\bibnamefont {Prasad}},\
  }\bibfield  {title} {\bibinfo {title} {Quantum limited superresolution of an
  incoherent source pair in three dimensions},\ }\href
  {https://doi.org/10.1103/PhysRevLett.121.180504} {\bibfield  {journal}
  {\bibinfo  {journal} {Phys. Rev. Lett.}\ }\textbf {\bibinfo {volume} {121}},\
  \bibinfo {pages} {180504} (\bibinfo {year} {2018})}\BibitemShut {NoStop}%
\bibitem [{\citenamefont {\ifmmode \check{R}\else
  \v{R}\fi{}eh\'a\ifmmode~\check{c}\else \v{c}\fi{}ek}\ \emph
  {et~al.}(2018)\citenamefont {\ifmmode \check{R}\else
  \v{R}\fi{}eh\'a\ifmmode~\check{c}\else \v{c}\fi{}ek}, \citenamefont {Hradil},
  \citenamefont {Koutn\'y}, \citenamefont {Grover}, \citenamefont {Krzic},\
  and\ \citenamefont {S\'anchez-Soto}}]{rehacek2018}%
  \BibitemOpen
  \bibfield  {author} {\bibinfo {author} {\bibfnamefont {J.}~\bibnamefont
  {\ifmmode \check{R}\else \v{R}\fi{}eh\'a\ifmmode~\check{c}\else
  \v{c}\fi{}ek}}, \bibinfo {author} {\bibfnamefont {Z.}~\bibnamefont {Hradil}},
  \bibinfo {author} {\bibfnamefont {D.}~\bibnamefont {Koutn\'y}}, \bibinfo
  {author} {\bibfnamefont {J.}~\bibnamefont {Grover}}, \bibinfo {author}
  {\bibfnamefont {A.}~\bibnamefont {Krzic}},\ and\ \bibinfo {author}
  {\bibfnamefont {L.~L.}\ \bibnamefont {S\'anchez-Soto}},\ }\bibfield  {title}
  {\bibinfo {title} {Optimal measurements for quantum spatial
  superresolution},\ }\href {https://doi.org/10.1103/PhysRevA.98.012103}
  {\bibfield  {journal} {\bibinfo  {journal} {Phys. Rev. A}\ }\textbf {\bibinfo
  {volume} {98}},\ \bibinfo {pages} {012103} (\bibinfo {year}
  {2018})}\BibitemShut {NoStop}%
\bibitem [{\citenamefont {Grace}\ \emph {et~al.}(2020)\citenamefont {Grace},
  \citenamefont {Dutton}, \citenamefont {Ashok},\ and\ \citenamefont
  {Guha}}]{Grace2020c}%
  \BibitemOpen
  \bibfield  {author} {\bibinfo {author} {\bibfnamefont {M.~R.}\ \bibnamefont
  {Grace}}, \bibinfo {author} {\bibfnamefont {Z.}~\bibnamefont {Dutton}},
  \bibinfo {author} {\bibfnamefont {A.}~\bibnamefont {Ashok}},\ and\ \bibinfo
  {author} {\bibfnamefont {S.}~\bibnamefont {Guha}},\ }\bibfield  {title}
  {\bibinfo {title} {{Approaching quantum-limited imaging resolution without
  prior knowledge of the object location}},\ }\href@noop {} {\bibfield
  {journal} {\bibinfo  {journal} {Journal of the Optical Society of America A}\
  }\textbf {\bibinfo {volume} {37}},\ \bibinfo {pages} {1288} (\bibinfo {year}
  {2020})}\BibitemShut {NoStop}%
\bibitem [{\citenamefont {Goodman}(1985)}]{Goo85Statistical}%
  \BibitemOpen
  \bibfield  {author} {\bibinfo {author} {\bibfnamefont {J.~W.}\ \bibnamefont
  {Goodman}},\ }\href@noop {} {\emph {\bibinfo {title} {Statistical Optics}}}\
  (\bibinfo  {publisher} {John Wiley \& Sons},\ \bibinfo {year}
  {1985})\BibitemShut {NoStop}%
\bibitem [{\citenamefont {Mandel}\ and\ \citenamefont {Wolf}(1995)}]{MW95}%
  \BibitemOpen
  \bibfield  {author} {\bibinfo {author} {\bibfnamefont {L.}~\bibnamefont
  {Mandel}}\ and\ \bibinfo {author} {\bibfnamefont {E.}~\bibnamefont {Wolf}},\
  }\href@noop {} {\emph {\bibinfo {title} {Optical Coherence and Quantum
  Optics}}}\ (\bibinfo  {publisher} {Cambridge University Press, Cambridge},\
  \bibinfo {year} {1995})\BibitemShut {NoStop}%
\bibitem [{\citenamefont {Labeyrie}\ \emph {et~al.}(2006)\citenamefont
  {Labeyrie}, \citenamefont {Lipson},\ and\ \citenamefont
  {Nisenson}}]{Labeyrie2006}%
  \BibitemOpen
  \bibfield  {author} {\bibinfo {author} {\bibfnamefont {A.}~\bibnamefont
  {Labeyrie}}, \bibinfo {author} {\bibfnamefont {S.~G.}\ \bibnamefont
  {Lipson}},\ and\ \bibinfo {author} {\bibfnamefont {P.}~\bibnamefont
  {Nisenson}},\ }\href@noop {} {\emph {\bibinfo {title} {An introduction to
  optical stellar interferometry}}}\ (\bibinfo  {publisher} {Cambridge
  University Press},\ \bibinfo {year} {2006})\BibitemShut {NoStop}%
\bibitem [{\citenamefont {Gottesman}\ \emph {et~al.}(2012)\citenamefont
  {Gottesman}, \citenamefont {Jennewein},\ and\ \citenamefont
  {Croke}}]{Gottesman2012}%
  \BibitemOpen
  \bibfield  {author} {\bibinfo {author} {\bibfnamefont {D.}~\bibnamefont
  {Gottesman}}, \bibinfo {author} {\bibfnamefont {T.}~\bibnamefont
  {Jennewein}},\ and\ \bibinfo {author} {\bibfnamefont {S.}~\bibnamefont
  {Croke}},\ }\bibfield  {title} {\bibinfo {title} {Longer-baseline telescopes
  using quantum repeaters},\ }\href
  {https://doi.org/10.1103/PhysRevLett.109.070503} {\bibfield  {journal}
  {\bibinfo  {journal} {Phys. Rev. Lett.}\ }\textbf {\bibinfo {volume} {109}},\
  \bibinfo {pages} {070503} (\bibinfo {year} {2012})}\BibitemShut {NoStop}%
\bibitem [{\citenamefont {Tsang}(2011)}]{Tsang2011}%
  \BibitemOpen
  \bibfield  {author} {\bibinfo {author} {\bibfnamefont {M.}~\bibnamefont
  {Tsang}},\ }\bibfield  {title} {\bibinfo {title} {Quantum nonlocality in
  weak-thermal-light interferometry},\ }\href
  {https://doi.org/10.1103/PhysRevLett.107.270402} {\bibfield  {journal}
  {\bibinfo  {journal} {Phys. Rev. Lett.}\ }\textbf {\bibinfo {volume} {107}},\
  \bibinfo {pages} {270402} (\bibinfo {year} {2011})}\BibitemShut {NoStop}%
\bibitem [{\citenamefont {Ram}\ \emph {et~al.}(2006)\citenamefont {Ram},
  \citenamefont {Ward},\ and\ \citenamefont {Ober}}]{Ram2006}%
  \BibitemOpen
  \bibfield  {author} {\bibinfo {author} {\bibfnamefont {S.}~\bibnamefont
  {Ram}}, \bibinfo {author} {\bibfnamefont {E.~S.}\ \bibnamefont {Ward}},\ and\
  \bibinfo {author} {\bibfnamefont {R.~J.}\ \bibnamefont {Ober}},\ }\bibfield
  {title} {\bibinfo {title} {Beyond rayleigh's criterion: A resolution measure
  with application to single-molecule microscopy},\ }\href
  {https://doi.org/10.1073/pnas.0508047103} {\bibfield  {journal} {\bibinfo
  {journal} {Proc. Natl. Acad. Sci. U.S.A.}\ }\textbf {\bibinfo {volume}
  {103}},\ \bibinfo {pages} {4457} (\bibinfo {year} {2006})}\BibitemShut
  {NoStop}%
\bibitem [{\citenamefont {Pawley}(2006)}]{pawley2006}%
  \BibitemOpen
  \bibfield  {author} {\bibinfo {author} {\bibfnamefont {J.}~\bibnamefont
  {Pawley}},\ }\href@noop {} {\emph {\bibinfo {title} {Handbook of biological
  confocal microscopy}}},\ Vol.\ \bibinfo {volume} {236}\ (\bibinfo
  {publisher} {Springer Science \& Business Media},\ \bibinfo {year}
  {2006})\BibitemShut {NoStop}%
\bibitem [{\citenamefont {Zmuidzinas}(2003)}]{Zmuidzinas2003CramrRaoSL}%
  \BibitemOpen
  \bibfield  {author} {\bibinfo {author} {\bibfnamefont {J.}~\bibnamefont
  {Zmuidzinas}},\ }\bibfield  {title} {\bibinfo {title} {Cram{\'e}r-rao
  sensitivity limits for astronomical instruments: implications for
  interferometer design.},\ }\href@noop {} {\bibfield  {journal} {\bibinfo
  {journal} {Journal of the Optical Society of America. A, Optics, image
  science, and vision}\ }\textbf {\bibinfo {volume} {20 2}},\ \bibinfo {pages}
  {218} (\bibinfo {year} {2003})}\BibitemShut {NoStop}%
\bibitem [{\citenamefont {Yariv}\ and\ \citenamefont {Yeh}(2006)}]{Yariv2006}%
  \BibitemOpen
  \bibfield  {author} {\bibinfo {author} {\bibfnamefont {A.}~\bibnamefont
  {Yariv}}\ and\ \bibinfo {author} {\bibfnamefont {P.}~\bibnamefont {Yeh}},\
  }\href@noop {} {\emph {\bibinfo {title} {Photonics: Optical Electronics in
  Modern Communications}}},\ \bibinfo {edition} {6th}\ ed.\ (\bibinfo
  {publisher} {Oxford University Press},\ \bibinfo {year} {2006})\BibitemShut
  {NoStop}%
\bibitem [{\citenamefont {Braunstein}\ and\ \citenamefont
  {Caves}(1994)}]{PhysRevLett.72.3439}%
  \BibitemOpen
  \bibfield  {author} {\bibinfo {author} {\bibfnamefont {S.~L.}\ \bibnamefont
  {Braunstein}}\ and\ \bibinfo {author} {\bibfnamefont {C.~M.}\ \bibnamefont
  {Caves}},\ }\bibfield  {title} {\bibinfo {title} {Statistical distance and
  the geometry of quantum states},\ }\href
  {https://doi.org/10.1103/PhysRevLett.72.3439} {\bibfield  {journal} {\bibinfo
   {journal} {Phys. Rev. Lett.}\ }\textbf {\bibinfo {volume} {72}},\ \bibinfo
  {pages} {3439} (\bibinfo {year} {1994})}\BibitemShut {NoStop}%
\bibitem [{\citenamefont {Barndorff-Nielsen}\ and\ \citenamefont
  {Gill}(2000)}]{Barndorff_Nielsen_2000}%
  \BibitemOpen
  \bibfield  {author} {\bibinfo {author} {\bibfnamefont {O.~E.}\ \bibnamefont
  {Barndorff-Nielsen}}\ and\ \bibinfo {author} {\bibfnamefont {R.~D.}\
  \bibnamefont {Gill}},\ }\bibfield  {title} {\bibinfo {title} {Fisher
  information in quantum statistics},\ }\href
  {https://doi.org/10.1088/0305-4470/33/24/306} {\bibfield  {journal} {\bibinfo
   {journal} {Journal of Physics A: Mathematical and General}\ }\textbf
  {\bibinfo {volume} {33}},\ \bibinfo {pages} {4481} (\bibinfo {year}
  {2000})}\BibitemShut {NoStop}%
\bibitem [{\citenamefont {G.~A.~Paris}(2008)}]{Paris2008}%
  \BibitemOpen
  \bibfield  {author} {\bibinfo {author} {\bibfnamefont {M.}~\bibnamefont
  {G.~A.~Paris}},\ }\bibfield  {title} {\bibinfo {title} {Quantum estimation
  for quantum technology},\ }\href {https://doi.org/10.1142/S0219749909004839}
  {\bibfield  {journal} {\bibinfo  {journal} {International Journal of Quantum
  Information}\ }\textbf {\bibinfo {volume} {7}} (\bibinfo {year}
  {2008})}\BibitemShut {NoStop}%
\bibitem [{Note1()}]{Note1}%
  \BibitemOpen
  \bibinfo {note} {The authors credit Saikat Guha and Ranjith Nair for
  realizing this well-known fact, in the context of this problem}\BibitemShut
  {NoStop}%
\bibitem [{\citenamefont {Lupo}\ \emph {et~al.}(2020)\citenamefont {Lupo},
  \citenamefont {Huang},\ and\ \citenamefont {Kok}}]{Cosmo2020}%
  \BibitemOpen
  \bibfield  {author} {\bibinfo {author} {\bibfnamefont {C.}~\bibnamefont
  {Lupo}}, \bibinfo {author} {\bibfnamefont {Z.}~\bibnamefont {Huang}},\ and\
  \bibinfo {author} {\bibfnamefont {P.}~\bibnamefont {Kok}},\ }\bibfield
  {title} {\bibinfo {title} {Quantum limits to incoherent imaging are achieved
  by linear interferometry},\ }\href
  {https://doi.org/10.1103/PhysRevLett.124.080503} {\bibfield  {journal}
  {\bibinfo  {journal} {Phys. Rev. Lett.}\ }\textbf {\bibinfo {volume} {124}},\
  \bibinfo {pages} {080503} (\bibinfo {year} {2020})}\BibitemShut {NoStop}%
\bibitem [{\citenamefont {{Hayashi}}\ and\ \citenamefont
  {{Matsumoto}}(2005)}]{2005atqs.book162H}%
  \BibitemOpen
  \bibfield  {author} {\bibinfo {author} {\bibfnamefont {M.}~\bibnamefont
  {{Hayashi}}}\ and\ \bibinfo {author} {\bibfnamefont {K.}~\bibnamefont
  {{Matsumoto}}},\ }\bibinfo {title} {{Statistical Model with Measurement
  Degree of Freedom and Quantum Physics}},\ in\ \href
  {https://doi.org/10.1142/9789812563071_0014} {\emph {\bibinfo {booktitle}
  {Asymptotic Theory of Quantum Statistical Inference: Selected Papers.~Edited
  by HAYASHI MASAHITO.~Published by World Scientific Publishing Co.~Pte.~Ltd.,
  2005.~ISBN \#9789812563071, pp.~162-169}}},\ \bibinfo {editor} {edited by\
  \bibinfo {editor} {\bibfnamefont {M.}~\bibnamefont {{Hayashi}}}}\ (\bibinfo
  {publisher} {World Scientific Publishing Co},\ \bibinfo {year} {2005})\ pp.\
  \bibinfo {pages} {162--169}\BibitemShut {NoStop}%
\bibitem [{\citenamefont {Gill}\ and\ \citenamefont
  {Massar}(2000)}]{PhysRevA.61.042312}%
  \BibitemOpen
  \bibfield  {author} {\bibinfo {author} {\bibfnamefont {R.~D.}\ \bibnamefont
  {Gill}}\ and\ \bibinfo {author} {\bibfnamefont {S.}~\bibnamefont {Massar}},\
  }\bibfield  {title} {\bibinfo {title} {State estimation for large
  ensembles},\ }\href {https://doi.org/10.1103/PhysRevA.61.042312} {\bibfield
  {journal} {\bibinfo  {journal} {Phys. Rev. A}\ }\textbf {\bibinfo {volume}
  {61}},\ \bibinfo {pages} {042312} (\bibinfo {year} {2000})}\BibitemShut
  {NoStop}%
\bibitem [{\citenamefont {Hayashi}(2006)}]{Hayashi2006}%
  \BibitemOpen
  \bibfield  {author} {\bibinfo {author} {\bibfnamefont {M.}~\bibnamefont
  {Hayashi}},\ }\href@noop {} {\emph {\bibinfo {title} {Quantum Information: An
  Introduction}}},\ \bibinfo {edition} {1st}\ ed.\ (\bibinfo  {publisher}
  {Springer-Verlag, Berlin Heidelberg},\ \bibinfo {year} {2006})\BibitemShut
  {NoStop}%
\bibitem [{\citenamefont {Fujiwara}(2006)}]{Fujiwara_2006}%
  \BibitemOpen
  \bibfield  {author} {\bibinfo {author} {\bibfnamefont {A.}~\bibnamefont
  {Fujiwara}},\ }\bibfield  {title} {\bibinfo {title} {Strong consistency and
  asymptotic efficiency for adaptive quantum estimation problems},\ }\href
  {https://doi.org/10.1088/0305-4470/39/40/014} {\bibfield  {journal} {\bibinfo
   {journal} {Journal of Physics A: Mathematical and General}\ }\textbf
  {\bibinfo {volume} {39}},\ \bibinfo {pages} {12489} (\bibinfo {year}
  {2006})}\BibitemShut {NoStop}%
\end{thebibliography}%

\end{document}